\documentclass[10pt,journal,compsoc]{IEEEtran}

\usepackage{caption}
\usepackage{adjustbox}
\usepackage{amsmath}
\usepackage{xspace}
\usepackage{soul}
\usepackage{diagbox}
\usepackage{multirow}
\usepackage{url}
\usepackage{booktabs}
\usepackage{xcolor}

\usepackage{algorithmicx}
\usepackage[ruled,vlined,linesnumbered]{algorithm2e}
\usepackage{setspace}

\SetKwInput{KwInput}{Input}                
\SetKwInput{KwOutput}{Output}              
\SetKwInput{KwRequire}{Require}            

\graphicspath{{../}{Figure/}}
\newcommand{\parahead}[1]{\noindent %
	{\bfseries #1.}}

\newcommand{\ie}{\emph{i.e.}\xspace}
\newcommand{\eg}{\emph{e.g.}\xspace}

\newcommand{\systemname}{DVFSLM\xspace}
\newcommand{\slm}{SLM\xspace}
\newcommand{\slms}{SLMs\xspace}
\newcommand{\dvfs}{DVFS\xspace}

\title{DVFS for Small Language Model Inference on Mobile Edge Devices}

\begin{document}
\IEEEtitleabstractindextext{
\author{Jiesong Chen, Lixiang Han, Jiani Cao, Zhaoxi Yue, ~\IEEEmembership{Member,~IEEE}, and Zhenjiang Li,~\IEEEmembership{Senior Member,~IEEE}
}

\IEEEcompsocitemizethanks{\IEEEcompsocthanksitem J.~Chen, L.~Han, J.~Cao, Z.~Yue and Z.~Li are with the Department of Computer Science, City University of Hong Kong, Hong Kong, China. E-mail: jiesochen2-c@my.cityu.edu.hk, lxhan2-c@my.cityu.edu.hk, jncao2-c@my.cityu.edu.hk, zhaoxiyue2-c@my.cityu.edu.hk, zhenjiang.li@cityu.edu.hk}
\IEEEcompsocitemizethanks{\IEEEcompsocthanksitem Corresponding author: Zhenjiang Li}

\begin{abstract}
This paper presents DVFSLM, a new dynamic voltage and frequency scaling (DVFS) design for energy-efficient inference of small language models (SLMs) on mobile edge devices. The growing demand for local execution of language models has driven the adoption of SLMs, which balance computational feasibility with good inference performance. However, energy efficiency remains a critical challenge, since even miniaturized SLMs impose significant energy consumption, impacting application quality, device reliability, and environmental sustainability. Existing DVFS solutions, designed for cloud-based large models or generic mobile workloads, fail to address the unique workload characteristics of SLMs, resulting in wasted energy or excessive latency. Unlike prior work, DVFSLM explicitly addresses two key challenges: 1) the complex interdependencies of processor frequencies, power and latency across autoregressive token generations, and 2) hardware opacity, where the individual power and latency contributions from different processors (GPU, CPU and EMC) are obscured during collaborative execution. To address these, DVFSLM introduces workload-aware power and latency estimators that analyze core matrix operations and correlate them with hardware metadata, enabling precise estimations of how frequency adjustments impact power and latency. These estimations drive a runtime DVFS governor that coordinates the GPU and EMC frequencies with a profiled CPU-frequency threshold, minimizing the energy per token while satisfying configurable token-generation deadlines. Extensive experiments on a rich set of SLMs show that DVFSLM reduces the energy per token by up to 12.4\% over the latest built-in governors and up to 8.4\% over the state-of-the-art GearDVFS, while improving the latency quality of service (QoS) by up to 93.12\% and 69.14\%, respectively.
\end{abstract}
}

\maketitle

\section{Introduction}
\label{sec:intro}

As large language models increasingly benefit various aspects of our lives, there is a growing demand to embed the capability of language model inference directly into mobile edge devices for \textit{local execution}, such as on the NVIDIA Jetson series and other systems-on-chips~\cite{slmonjetson,song2023powerinfer}. Pushing this capability to the mobile edge is reshaping the way future smart-city applications will be developed, for example, language models will automate next-generation controllers in autonomous machines like vehicles and robots~\cite{liang2023code,huang2025modality}, enhance devices with language model-assisted kernels~\cite{yuan2024mobile,wen2024autodroid}, enable Internet of Things (IoT) management through language-based interactions~\cite{king2024sasha}, and more.

This trend is inevitable~\cite{yin2024llm}, due to the inherent deficiencies of existing cloud-based large language models (LLMs) that hinder their effectiveness to support these applications, including \textit{1) latency instability} (connection quality or even connectivity can be easily affected by various factors, especially over wireless~\cite{liu2024andes, kong2023accumo}) and \textit{2) data privacy concerns}. Fortunately, recent breakthroughs in designing Small versions of Language Models (\textbf{SLMs})~\cite{javaheripi2023phi,zhang2022opt} have made the local execution of language models increasingly feasible. \slms have much fewer parameters (\eg, TinyLLaMA and MobileLLM have only a few billion parameters, which is 120--1400x fewer than GPT-4) while still achieving excellent inference performance, and thus have great potential to enable practical applications or services~\cite{shen2025gpiot}.

However, unlike traditional tasks, even a miniaturized \slm remains computationally intensive~\cite{lu2024small}. When it runs locally, the \textit{energy consumption} of the device becomes a critical problem, requiring \textit{energy-efficient \slm inference} via effective \textit{power governing} with the following significance: 

\textbf{1) Environmental consideration.} Local SLM inference can add nontrivial incremental energy use to mobile edge devices. For instance, even if a device runs an SLM for only one hour per day, the annual electricity use is on the order of 6.2–9.5 kWh~\cite{jetsonbenchmarks}. Given the large number of mobile edge devices globally, \eg, 7.5 billion~\cite{worldpopulation}, the improvement in energy efficiency could translate into meaningful electricity savings and associated emission reductions~\cite{lai2014electricity}.

\textbf{2) Performance and reliability.} In addition to energy conservation, the significance of power governing for \slms on mobile edge devices is also about reducing heat generation, as the compact device architecture makes heat dissipation difficult~\cite{kim2021ztt}. Power governing ensures \slms to fully exploit the inherent computing capability of the device without overheating to affect the performance~\cite{kim2021ztt}. Continuous high energy consumption further affects hardware life cycle~\cite{ratkovic2015overview}, compromising the lifespan of the device and the reliability of \slm-enabled applications.

Given that dynamic voltage and frequency scaling (\dvfs) is the primary power-governing technology for mobile edge devices, we aim to design a new and effective \dvfs for \slms in this paper. The principle of \dvfs is to adjust the operating frequencies of processors, GPU, CPU and EMC (External Memory Controller), at runtime (also known as \textbf{frequency scaling}) to match the energy consumption of those processors with the workload~\cite{kim2018survey}. In other words, \dvfs targets to allocate \textit{reasonable} energy (without waste) to power processors to complete the workload on time~\cite{kim2021ztt}, but the hurdle for \slms is the sophisticated relation between the complex workloads and the matched computing power.

While the high-end platforms (\eg, clouds) have their own \dvfs, it is designed to handle large-scale workloads (\eg, for LLMs) distributed across GPU clusters~\cite{gu2014optimal,zhang2024improving}, which focus on batch size adjustment~\cite{kakolyris2025throttll} and GPU task scheduling~\cite{stojkovic2025dynamollm}. Frequency scaling is also performed but at intervals of several seconds~\cite{stojkovic2025dynamollm}. Since mobile edge devices have much lower computing power and use a single GPU~\cite{lu2024small}, the dominance of \slm workloads on mobile edge devices requires fine-grained frequency scaling during \slm execution, typically at sub-second levels~\cite{lin2023workload,kim2021ztt}, to adapt to the workload dynamics. Therefore, cloud's \dvfs is unsuitable. Existing \dvfs on mobile edges can achieve sub-second operation~\cite{farazmand2018dynamic}, but lack effective designs (detailed in Section~\ref{sec:bg}) due to the unique challenges posed by \slms:

\textbf{1) Autoregressive workload.} Unlike traditional models, \slms usually have long inference times (up to several seconds) since \slm's complexity leads to \textit{massive} computations to generate each token, and its autoregressive nature further requires running \slm \textit{repeatedly}. Inefficient frequency scaling can lead to latency \textit{accumulation} over autoregressions to unacceptable levels (conservative scaling) or minimal energy reduction (aggressive scaling). Thus, the first challenge lies in the unknown relationship for the interdependent triplet of frequencies, power, and latency, which needs to be characterized in order to optimize \slms' energy consumption\footnote{\textbf{Power consumption} refers to the instantaneous rate at which a device consumes energy, while \textbf{energy consumption} is the amount of energy used over a period of time. \dvfs controls power consumption by adjusting processor frequencies, thereby reducing the energy consumption of a device.} without sacrificing the latency requirement.

\textbf{2) Estimator design.} Autoregression also causes a cumulative workload as each output token is fed back to generate the next one (\ie, the varying of context length). The relationship between frequencies and power and latency must account for such workload dynamics. Even if such a relationship can be derived, the development of estimators for quantifying power and latency to select the optimal frequencies further faces the problem of hardware opacity: When processors (GPU, CPU and EMC) collaborate for \slm inference, the device only exposes the overall power and latency performance of inference. It masks the contribution of each processor, which however is necessary for determining the parameters of each estimator.

In this paper, we propose a new system called \systemname to address these challenges. Since \slm's core computations can be expressed as a series of matrix multiplications~\cite{vaswani2017attention}, we find that the impact of autoregression on the workload can be unified and reflected through the scale of these multiplications, and the device hardware data~\cite{mazzola2022data} widely adopted in prior power governing has great potential for achieving workload-aware \slm power and latency estimations, but underexplored. Based on this, we 1) introduce a formulation to capture the power and latency for the upcoming \slm workload across a range of operating frequencies, 2) solve the hardware opacity issue to design two estimators that can estimate the power and latency of running that workload using different frequency combinations, and 3) utilize these two estimators to drive the governor, which periodically adjusts processor frequencies to minimize energy consumption while meeting the required token generation deadline (see Section~\ref{sec:bg}). We further study the impact of CPU frequency scaling and find that a profiled CPU-frequency threshold is sufficient to avoid GPU stalls while keeping the CPU energy low. Therefore, \systemname is designed as a GPU--EMC coordinated \dvfs governor with a profiled CPU-frequency threshold.

We implement a prototype of \systemname and conduct comprehensive evaluations on NVIDIA Jetson AGX Orin and Orin NX using prominent \slms: GPT2~\cite{radford2019language}, Qwen~\cite{qwen2}, TinyLLaMA~\cite{zhang2024tinyllama} and OLMoE~\cite{muennighoff2024olmoe}, including both classical and mixture of experts (MoE) structures. Experimental results show that \systemname outperforms the latest built-in \dvfs governors on mobile edge devices and the state-of-the-art GearDVFS~\cite{lin2023workload}, improving the energy efficiency by up to 12.4\% and 8.4\%, respectively. Evaluations also reveal that existing solutions fail to meet token generation deadlines, \ie, quality of service (QoS) requirements, exhibiting 10.9--45.8\% QoS degradation. In contrast, \systemname simultaneously maximizes energy efficiency and guarantees latency compliance, improving QoS by up to 93.12\% compared to built-in governors and up to 69.14\% compared to GearDVFS in the experiments. In addition, a series of micro-benchmark evaluations confirm \systemname's robustness under different scenarios. Finally, \systemname itself is lightweight, consuming only 0.04 W of power.

Although edge devices such as the NVIDIA Jetson series are often evaluated under plugged-in settings in research, they are widely deployed in power-constrained edge systems, including UAVs~\cite{nvidia_aerialtronics_jetson_uav}, mobile robots~\cite{nvidia_serve_robotics_jetson}, and autonomous vehicles~\cite{bhatt2024watonobus}, where energy efficiency determines operational endurance and thermal headroom. In these scenarios, reducing the energy consumed per generated token not only extends battery lifetime, but also lowers heat generation and sustains throughput under tight power and thermal budgets, benefiting both battery-powered and plugged-in edge deployments.

The core contributions of this paper are summarized as follows:

\begin{itemize}
    \item We formulate a \textit{deadline-aware} energy minimization problem for autoregressive \slm decoding on mobile edge devices, which minimizes the energy per token while satisfying a configurable token generation deadline. The deadline does not need to be known in advance and can be adjusted at any time.
    \item We address the hardware opacity challenge by designing workload- and metadata-aware power and latency estimators. They convert the upcoming token workload into the scale of matrix operations and correlate it with hardware metadata, enabling accurate estimation of how frequency adjustments affect power and latency.
    \item We develop a lightweight runtime governor that coordinates the GPU and EMC frequencies with a profiled CPU-frequency threshold, and build a prototype system. Extensive experiments demonstrate significant gains over the latest built-in \dvfs governors on commercial mobile edge devices and the state-of-the-art learning-based design.
\end{itemize}

\section{Background and Motivation}
\label{sec:bg}

\subsection{Energy Saving Opportunities for \slms}

\begin{figure}[t]
\centering
    \includegraphics[width=0.9\linewidth]{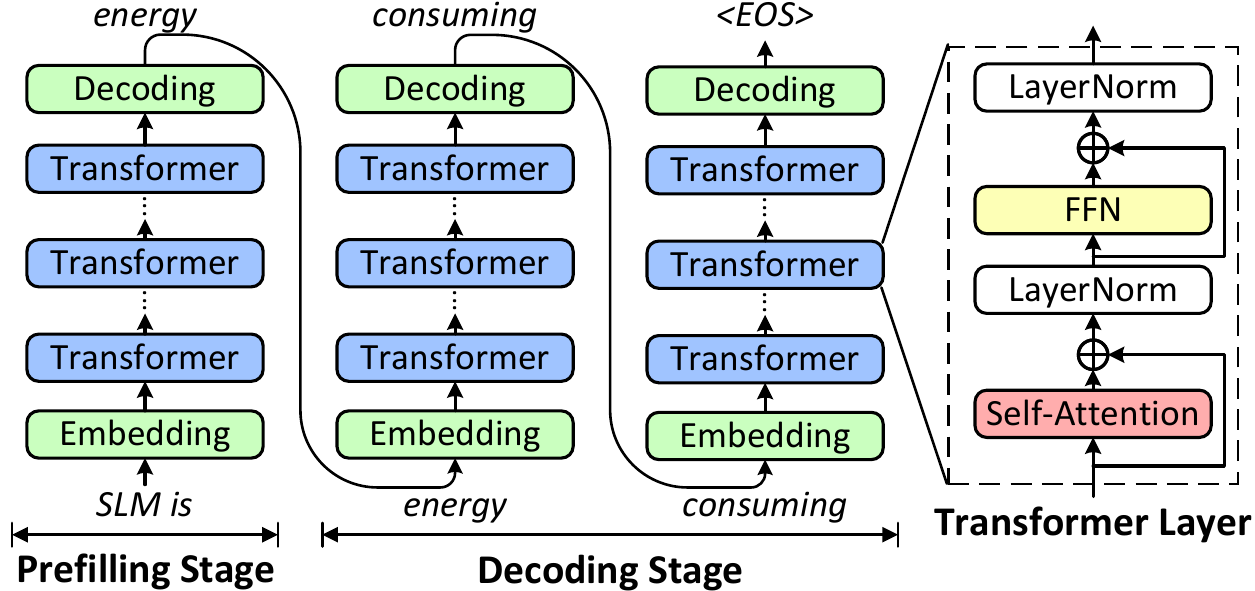}
    \caption{Illustration of autoregressive inference of \slms.}
    \label{fig:slm}
    \vspace{-.1in}
\end{figure}

\parahead{1) Workload Characteristics of \slms} \slms typically employ 12--48 transformer layers~\cite{zhang2022opt,javaheripi2023phi}, each containing a series of processing blocks such as self-attention and feed-forward networks (FFNs), as illustrated in Figure~\ref{fig:slm}. 

\slm inference occurs in two stages: an initial \textit{prefilling} stage that processes the input prompt, followed by the primary \textit{decoding} stage that generates output tokens iteratively. During decoding, each newly generated token is appended to the input sequence, requiring querying for all previous tokens through self-attention. This \textbf{autoregressive} nature results in progressively increasing computational workload per token throughout the decoding stage of the inference.


\parahead{2) Processors and \dvfs} On mobile edge devices, three processors, GPU, CPU and EMC, collaborate to complete SLM inference. Specifically, GPU performs the main computations, CPU manages general control flow, preprocesses inputs, and offloads compute-heavy operations to the GPU, and EMC ensures rapid data access for GPU and CPU.


For each processor, its power consumption $p \propto V^2f$, and the voltage $V$ is positively correlated with the operating frequency $f$~\cite{gu2014optimal,zhang2024improving}. Therefore, $p \propto f^3$. Each processor has discrete frequency levels to choose from, which brings the trade-off between energy consumption and processing latency. As a result, \dvfs periodically adjusts the $f$ of each processor to balance the energy-latency trade-off~\cite{kim2018survey}. 

\parahead{3) Energy Saving Opportunities} Due to autoregression, \slms cannot generate all output tokens at once, which complicates their workload but provides energy saving opportunities. In many \slm-based applications, the output tokens are \textbf{consumed} (read) by users, such as novel human-computer interaction~\cite{yang2024talk2care}, chatbots~\cite{kim2022soda}, text translation or summarization~\cite{hermann2015teaching, fabbri2019multi}, etc. In such cases, generating 15--20 tokens/s is usually sufficient, since humans usually read around 300 tokens per minute (5 tokens/s)~\cite{brysbaert2019many}. Assuming an aggressive \dvfs that increases the token generation rate, the improvement in service quality is negligible, but the energy consumption may increase significantly. 

Hence, \textbf{if output tokens can be generated in time before consumed, \dvfs does not have to increase processor frequencies, which can save substantial energy. When} \textbf{a user reads information much faster or the token consumer is another program or device~\cite{yang2024talk2care}, the rationale is the same, and only the \textbf{deadline} may become more urgent}. As this token-generation deadline needs to be fully configurable, effective \slm power governing should not be tied to any fixed token generation rate and needs to adjust flexibly when the deadline requirement is updated, enabling adaptation to evolving user or application needs.\footnote{Our design will provide the feasibility of adapting to various token generation rates (for example, \slm may not know the appropriate rate before deployment or the user may actively adjust it), but the method itself does not require active changing of the token generation rate.}

These opportunities apply to the primary decoding stage. Since the prefilling stage is significantly shorter than decoding and dictates the time-to-first-token, \dvfs should use high frequencies to minimize this initial latency, a critical factor for the user's perception of system responsiveness.

\begin{figure}[t]
    \begin{center}
        \includegraphics[width=.9\linewidth]
        {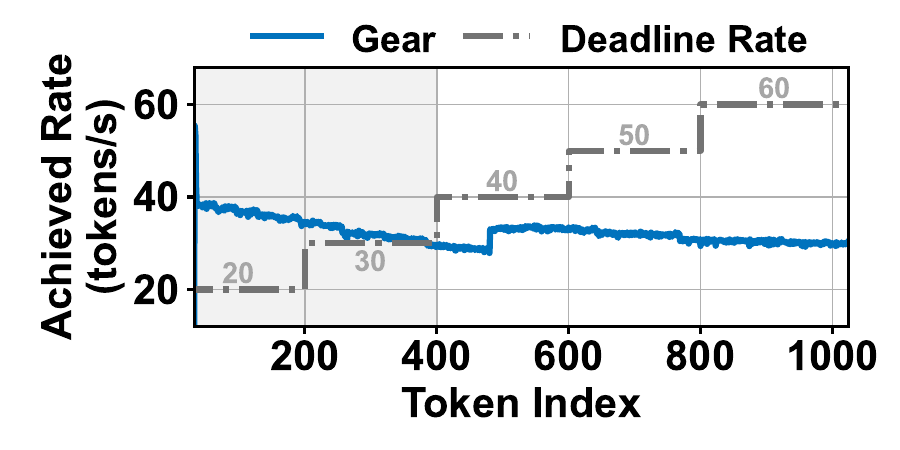}
    \end{center}
    \vspace{-.15in}
    \caption{We apply the latest GearDVFS to GPT2-large. We vary the deadline rates at runtime of running GPT2-large only for the sake of illustration. As revealed, GearDVFS lacks flexibility in complying with different rate requirements, which will severely limit its adoption in practice.}
    \label{fig:pre}
    \vspace{-.1in}
\end{figure}

\subsection{Why Existing DVFS is Inefficient for \slms?}

Before introducing our design, we first discuss why existing \dvfs on mobile edge devices underperforms for \slms.

A series of learning-based \dvfs methods, such as GearDVFS~\cite{lin2023workload} and zTT~\cite{kim2021ztt}, have been proposed to enhance energy efficiency for computationally intensive deep learning workloads on mobile edge devices. These methods identify a key indicator (device- or application-specific) and determine its target value to balance energy consumption and performance metrics (\eg, processor utilization~\cite{lin2023workload} or frame rate in video-based applications~\cite{kim2021ztt}). By dynamically adjusting frequencies to maintain the indicator close to its target value, they avoid explicitly modeling the intricate relationship between frequency, power, and performance.

However, \dvfs operates as a system-level service and each power governing needs to be completed in millisecond levels~\cite{kim2018survey}, limiting the complexity of the underlying learning models in the \dvfs governor. For \slms, which exhibit dynamic and sophisticated workloads, these lightweight models fail to capture the critical workload characteristics, leading to two inherent limitations. 

First, the token generation fails to fulfill the generation rate requirement, degrading user experience. To illustrate, we evaluate the state-of-the-art DVFS governor on mobile edge devices, \ie GearDVFS~\cite{lin2023workload}, on GPT2-large. Since GearDVFS originally optimizes only processor utilization and temperature rather than application-level latency, we adapt it to the \slm inference scenario for a fair comparison. The detailed implementation is described in Section~\ref{s:eval}. Figure~\ref{fig:pre} shows that when trained for 20 tokens/s but tested at various \textbf{deadline rates (minimum required rates, 20--60 tokens/s)}, it satisfies the rate requirement at low rate regions (20--30 tokens/s), \ie, the actual rate achieved is no less than the deadline rate as shown in the shaded area in the figure, but lags behind as deadline rates increase.\footnote{Initially, GearDVFS achieves a rate of about 40 tokens/s. Although it meets the deadline requirement (\eg, $>$ 20), it does not strictly follow the deadline rate. We defer the explanation to Figure ~\ref{fig:eval:adjust_target} when we have enough context.} Second, we will also see in later evaluation (Section~\ref{s:eval}) that no matter what rate it operates at, its energy consumption is still suboptimal.

We note that the staircase-like deadline pattern in Figure~\ref{fig:pre} is not intended to mimic a real-world usage pattern in which the required rate changes every short interval. Rather, it is a controlled stimulus that probes how GearDVFS responds to different deadline requirements. The key point is that the required rate may not be known before deployment and should be adjustable at runtime. The lagging behavior in Figure~\ref{fig:pre} is also not a random artifact: GearDVFS reacts to aggregate runtime signals (\eg, token latency, utilization, and temperature) instead of directly solving a deadline-constrained energy minimization problem, so its reaction is governed by the dynamics of these signals rather than by the deadline change itself.

In addition to these learning-based methods, the latest built-in governors on mobile edges are threshold-based (Section~\ref{s:related}) with similar inefficiencies, evaluated in Section~\ref{s:eval}. These limitations underscore the need for a \textbf{general} (target rate-agnostic), \textbf{effective} (capable of handling complex \slm workloads), and \textbf{lightweight} \dvfs tailored to \slms.
\section{System Design}
\label{s:design}

\begin{figure}[t]
    \begin{center}
        \includegraphics[width=.96\linewidth]{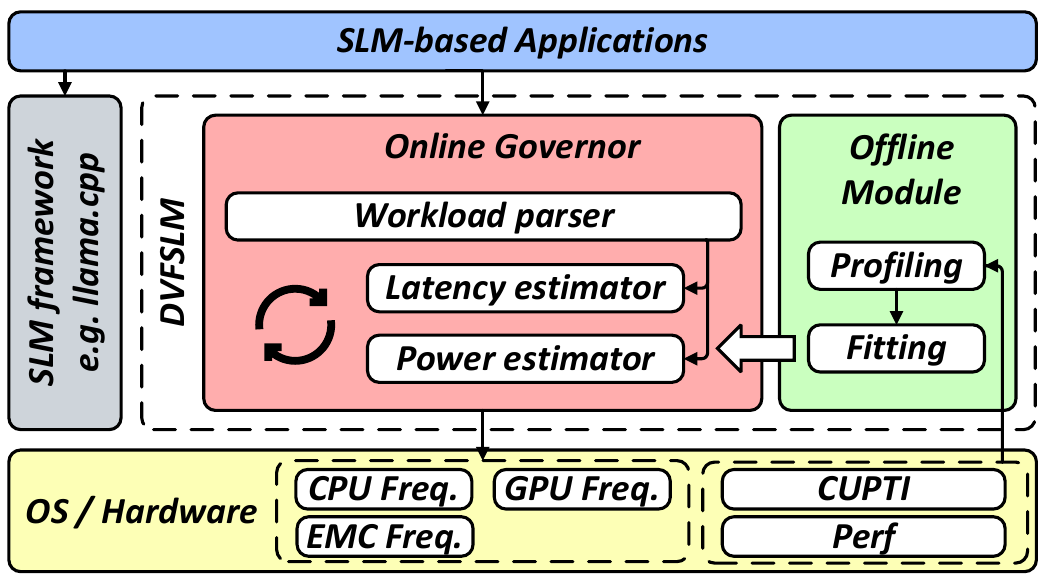}
    \end{center}
    \vspace{-.1in}
    \caption{Architecture of the \systemname design.}
    \label{fig:architecture}
    \vspace{-.15in}
\end{figure}

Figure~\ref{fig:architecture} illustrates the architecture of \systemname, a versatile middleware layer positioned between the OS kernel and applications, which requires no modifications to \slms and inference frameworks, making the design easy to deploy.

\textbf{1) Online governor.} The online governor derives the upcoming decoding workload and refers to the power and latency estimators to select the GPU and EMC frequencies that minimize the energy per token while meeting the deadline. The selected frequencies are the optimal ones within the searched GPU--EMC space conditioned on a critical CPU frequency $\hat{f}_c$. During the decoding stage, the governor scales frequencies once per decision window (50 tokens by default). For the prefilling stage, the governor uses high frequencies to prioritize minimizing the initial waiting time.

\textbf{2) Offline module.} This module employs a novel strategy to determine the parameters of power and latency estimators. Executed as a one-time setup for each \slm per device, this offline phase ensures the system is calibrated for subsequent inference.

\subsection{Formulating \slm's Power Governing}

During \slm inference, the \dvfs governor adjusts the frequency $f_{r}$ of each processor $r$ (GPU, CPU and EMC) to minimize the energy consumption $e_i(\{f_r\})$ of generating each output token $x_i$, subjected to the token generation time $t_i(\{f_r\})$ not exceeding the deadline $\delta$. This is formalized as:
\begin{equation} \label{eqn:opt}
    \min\nolimits_{\{f_r\}} e_i(\{f_r\}); \quad s.t.~t_i(\{f_r\}) \le \delta,
\end{equation}
where $e_i(\{f_r\})=p_i(\{f_r\}) \times t_i(\{f_r\})$, and $p_i(\{f_r\})$ is the power consumption of each processor at its operating frequency $f_r$. Due to the long execution time of \slm inference, we minimize the energy consumption $e_i(\{f_r\})$ instead of the instantaneous power consumption $p_i(\{f_r\})$ in Eqn.~(\ref{eqn:opt}). By default, \systemname sets the deadline $\delta$ to 50 ms (\ie, 20 tokens per second), which is suitable for human reading. However, this parameter can be changed at any time if a faster rate is needed in different applications. We formulate Eqn.~(\ref{eqn:opt}) as a latency-constrained energy minimization problem, because token latency is the primary QoS requirement in our target applications. Other system factors, such as device temperature and storage can also be integrated into the optimization problem by adding new constraints.

\textbf{Two Estimators.} Solving this formulation to determine the decision variables $\{f_r\}$ (where $r$ represents GPU, CPU, and EMC) requires a precise characterization of the relationship between processor frequencies $\{f_r\}$ and the power consumption $p(\{f_r\})$ and the latency $t(\{f_r\})$ during each token generation. Therefore, we need the following two estimators in \systemname:
\begin{itemize}
    \item \textbf{Power estimator $\mathrm{PowE}_i$}: $\{f_r\} \rightarrow p_{i}(\{f_r\})$, estimating the total power consumption of the three processors when generating each output token $x_i$.
    \item \textbf{Latency estimator $\mathrm{LatE}_i$}: $\{f_r\} \rightarrow t_{i}(\{f_r\})$, estimating the total latency to generate each token $x_i$.
\end{itemize}

Before introducing our design, we note two key points:
\begin{itemize}
    \item First, the design of these two estimators is related to $x_i$, because due to the autoregression nature, the computational workload to generate each output token $x_i$ will vary with different context lengths. Typically, for a given frequency setting, a longer context results in higher latency and greater power consumption. The estimator should account for the current context length when predicting latency and power. In addition, due to the variety of context lengths, it is difficult to replace these two estimators by measuring the performance of power consumption and latency in advance.
    \item Second, token $x_i$ is the token that is about to be generated (its generation has not yet been started), and the selected processor frequencies by \dvfs will be applied to the generation of this upcoming token. 
\end{itemize}

Therefore, we need to first design an effective method to obtain and quantify the upcoming workload for the next token, based on which the two estimators are designed to complete power and latency estimation across different processor frequencies under different context lengths.

\textbf{Governor Operations.} Using these two estimators, the \dvfs governor can evaluate the power and latency for each $\{f_r\}$ combination to generate each output token $x_i$, and then determine the frequencies that are optimal within the GPU--EMC space according to Eqn.~(\ref{eqn:opt}), which can be solved efficiently in around 1 ms each time (Section~\ref{s:eval}). Therefore, the next crucial step is to design these two estimators.

\subsection{Design of \systemname Estimators}

The estimator design of ``$\mathrm{PowE}_i$: $\{f_r\} \rightarrow p_{i}(\{f_r\})$'' and ``$\mathrm{LatE}_i$: $\{f_r\} \rightarrow t_{i}(\{f_r\})$'' is essentially a learning problem. A natural question is: can we use neural networks as estimators to directly learn each relationship? We tried popular models or networks, such as MLP and SVR, and found that it is difficult to directly capture the sophisticated relationship
using small models or networks. Their errors are large (we perform an ablation study in Section~\ref{s:eval}). However, if we increase their model size or complexity to improve accuracy, it will not be suitable for \dvfs because \dvfs is a system-level service whose size and execution time should be lightweight~\cite{kim2018survey}.

\subsubsection{Observation}

In \systemname, we achieve the estimator design based on the following key observation. Power consumption and latency are not mere outcomes when processors execute a workload. They are also \textit{visible signatures} of the processor operations, reflecting their internal characteristics and runtime behaviors. Fortunately, mobile edge devices unlock the visibility into processor operations through the \textbf{metadata} of hardware performance counters (HPCs)~\cite{mazzola2022data}, which capture granular details like instruction retirements, global load instructions, cache misses, and more, forming a rich fingerprint of processor characteristics and behaviors. 

We observe that if we can explicitly reveal and quantify the core relationship behind these characteristics and behaviors, and ground the estimator design in such a relationship, a lightweight model (\eg, a regression model) will suffice. Because once the underlying relationship is revealed, the estimator does not need to be trained to explore the relationship on its own, but only needs to quantify the (given) relationship (thus, a small model is sufficient). Therefore, we apply this insight and leverage the device metadata as the underlying signals through a two-stage design:
\begin{itemize}
    \item \textit{1) Workload representation:} We first design a parser to convert the workload of the next token generation into a unified representation based on metadata. The reason for using the parser instead of accessing the metadata at runtime is that this workload is upcoming and the corresponding metadata is not yet available. Thus, we only access the metadata to train the parser.
    \item \textit{2) Power and latency estimation:} During the power governing runtime, we use the workload represented by metadata from the parser and our designed estimators to perform the power and latency estimation.
\end{itemize}

Next we introduce the detailed design of each step.

\begin{figure}[t]
    \begin{center}
        \includegraphics[width=0.95\linewidth]{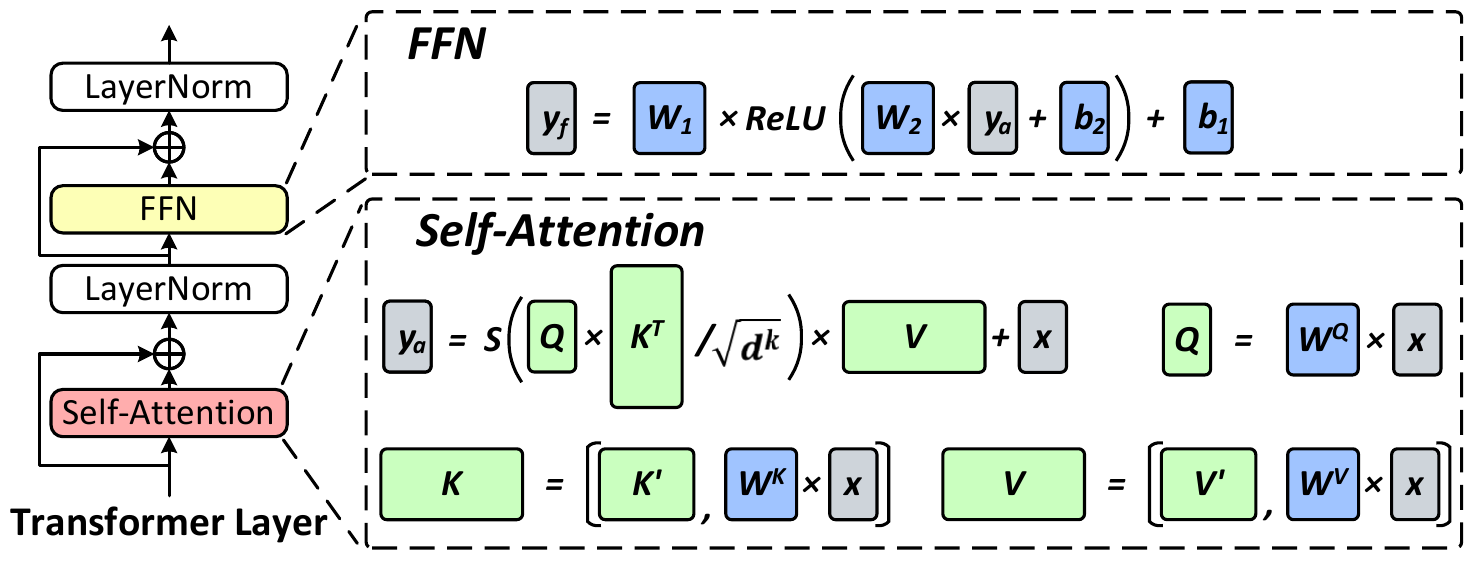}
    \end{center}
    \vspace{-.1in}
    \caption{Main calculations in the transformer layer.}
    \label{fig:slm_cal}
    \vspace{-.15in}
\end{figure}

\subsubsection{Workload Representation} For the newly generated token $x$, it will be used to generate the next token as follows. In each transformer layer, as shown in Figure~\ref{fig:slm_cal}, the calculation of the self-attention can be first expressed as:
\begin{equation}
    y_a(x)=S(QK^T/\sqrt{d^k})V + x,
\end{equation}
where $S(\cdot)$ is softmax, $Q = W^Qx$, $K=[K^\prime, W^Kx]$ and $V=[V^\prime, W^Vx]$. $W^Q$, $W^K$ and $W^V$ are the weights of attention, where $K$ and $V$ are concatenated to their previous-round values $K^\prime$ and $V^\prime$ (\ie, KV cache) to speed up the calculation. After attention, the layer norm and FFN are applied:
\begin{equation}
    y_f(y_a(x)) = W_1 \times ReLU(W_2 \times y_a(x) + b_2) + b_1,
\end{equation}
where $W_1$, $W_2$, $b_1$, and $b_2$ are parameters. Thus, the operation of a transformer layer can be expressed as $y_t(x) = y_f(y_a(x)) + y_a(x)$. The output is subsequently layer-normalized to the same dimension as $x$ and then passed to the next transformer layer. 

\textbf{\slm Workload-Awareness.} Therefore, if an \slm has $L$ transformer layers, the computation of any output token $x_i$ can be recursively expressed as follows:
\begin{equation} \label{eqn:slmrec}
    x_i = y_t(x_{i - 1}^L),~\text{where}~x_{i-1}^l = y_t(x_{i-1}^{l-1})~\text{and}~x_{i-1}^0 = x_{i-1},
\end{equation}
which is related to all previous tokens directly (\eg, $x_{i-1}$) or indirectly (\eg, other tokens
before $x_{i-1}$ from the KV cache). By extending the recursion of Eqn.~(\ref{eqn:slmrec}), the computation of $x_i$
can be expressed as a series of matrix multiplications:
\begin{equation} \label{eqn:expression}
    \{A_{i,j}^{m_{i,j}\times k_{i,j}} \times B_{i,j}^{k_{i,j}\times n_{i,j}}\}_{j=1}^I,
\end{equation}

where $m_{i,j}$, $k_{i,j}$, and $n_{i,j}$ denote the three dimensions of the $j$-th matrix multiplication required to generate token $x_i$. Specifically, $m_{i,j}$ and $n_{i,j}$ denote the two output dimensions, $k_{i,j}$ is the contracted inner dimension, and $I$ is the total number of matrix multiplications involved. Eqn.~(\ref{eqn:expression}) encapsulates the primary computations required for generating each output token $x_i$, and the \textbf{scale} of these matrices ($m_{i,j}$, $k_{i,j}$ and $n_{i,j}$) determines the volume of the computational workload. We observe that the scale, denoted as $\mathbf{s}$, can serve as a good workload fingerprint,
\begin{equation} \label{eqn:scale}
    \mathbf{s}_i = \{s_{i,j}\}_{j=1}^I, ~\text{where}~s_{i,j} = \langle m_{i,j}, k_{i,j}, n_{i,j} \rangle,
\end{equation}
and it could lead to an accurate conversion to the hardware metadata. With this capability, we can quantify metadata (needed by two estimators) about the workload that will happen (not yet) for the next output token.

\begin{figure}[t]
    \begin{center}
        \includegraphics[width=0.95\linewidth]{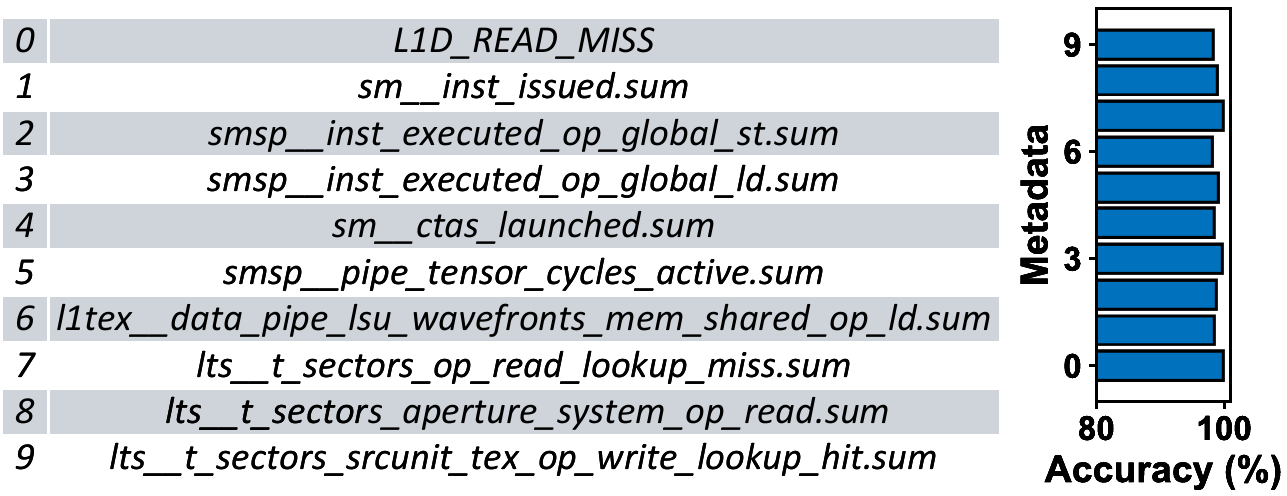}
    \end{center}
    \vspace{-.1in}
    \caption{(Left): Metadata selected for TinyLLaMA; (Right) Prediction accuracy for each metadata.}
    \label{fig:meta}
    \vspace{-.15in}
\end{figure}

\textbf{Parsing Scale into Metadata.} Mobile edge devices can expose rich HPC metadata~\cite{cuptievent}. The hardware metadata implicitly defines a latent hyper-space $\{\psi^h\}$, where each metadata $\psi^h$ represents a dimension. We find that the computational workload for an output token (in the form of scales) can be represented in this hyper-space by using even simple and lightweight methods, \eg, XGBoost~\cite{chen2016xgboost}. Using the full set of metadata is impractical due to high dimensionality and redundancy. We thus employ Pearson correlation coefficient to select a subset of metadata $\{\psi^h\}_{h=1}^M$ that is most relevant to the workload's characteristics (\eg, power and latency) and least correlated with each other as the input of the parser.

For instance, this process for TinyLLaMA yields $M = 10$ key metadata counters, shown in Figure 5 (Left). This subset creates a hardware fingerprint by capturing the most critical performance events during model inference. Specifically, metadata 1, 4, and 5 quantify the core computational load, reflecting instruction volume, parallelism, and Tensor Core utilization. Metadata 2 and 3 measure the main memory bandwidth by tracking global load and store operations. Metadata 0, 6--9 characterize the efficiency of the on-chip memory hierarchy, capturing  events from cache misses and hits to shared memory accesses. Factors such as KV-cache locality and memory fragmentation are implicitly captured in the selected metadata. Figure~\ref{fig:meta}(Right) further shows that the parser can use the scale of each output token computation to accurately predict the corresponding metadata, with an average accuracy of 99.89\%, laying a good foundation for subsequent power and latency estimation.

\textit{Summary.} To determine the processor frequencies for the next output token $x_i$, \systemname first derives the scale $\mathbf{s}_i$ of this upcoming computation from Eqn.~(\ref{eqn:slmrec}) to (\ref{eqn:scale}), and then parses the scale to the corresponding metadata as follows:
\begin{itemize}
    \item \textbf{Parser}: $\mathbf{s}_i \rightarrow \{\psi_i^h\}_{h=1}^M$,
\end{itemize}
where $\{\psi_i^h\}_{h=1}^M$ will act as the next step's estimation input.

\subsubsection{Power and Latency Estimation}

\systemname then needs to estimate the total power consumption and latency of generating each output token when using different frequency combinations, depending on GPU, CPU and EMC. Our design begins by examining each processor's behavior.

\textbf{Individual Processor Analysis.} For a given processor, its power consumption $p(f)$ is influenced by the following aspects~\cite{de2014energy}:
\begin{equation} \label{eqn:pf}
    p(f) = (1 + \phi) \eta C V^2 f,
\end{equation}
where parameters fall into two categories:
\begin{itemize}
    \item \textit{Hardware-dependent} parameters: 1) $\phi$: short-circuit power loss during logic gate transitions; 2) $C$: capacitance; 3) $V$: voltage; 4) $f$: operating frequency.
    \item \textit{Workload-dependent} parameters: $\eta$: fraction of the processor's transistors actively switching states during computation of the current workload.
\end{itemize}

Given the proportional relationship between $V$ and $f$~\cite{de2014energy, zhang2024improving, gu2014optimal}, we rephrase Eqn.~(\ref{eqn:pf}) to $p(f) = w_p \times f^3$, where $w_p = (1 + \phi) \eta C$. The cubic term is an empirical basis rather than a strict physical assumption. Jetson platforms may show non-cubic behavior in low-frequency regions. A more fine-grained piecewise regression or non-parametric model could potentially improve accuracy, but our current modeling strikes a balance between accuracy and simplicity. Using parsed workload metadata $\{\psi_i^h\}_{h=1}^M$, we find that even a lightweight regression model can effectively capture above hardware-dependent and workload-dependent impacts. As a pioneer attempt, our implementation employs linear regression to model $w_p$, which already achieves sufficient accuracy (validated in Section~\ref{s:eval}). Therefore, the power consumption for each processor $r$ can be expressed as:
\begin{equation} \label{eqn:ep}
    p_{i}(f_r) = (\lambda_r^0 + \sum\nolimits_{h=1}^{N_r} \lambda_r^h \psi_i^h) f_r^3,
\end{equation}
where $N_r$ denotes the regression order and $\{\lambda_r^h\}$ represent the regression parameters to be determined. The linear regression is applied on top of the physically-motivated frequency basis $f^3$, rather than treating power as a linear function of the raw frequency.

\begin{table}[t]
\centering

\caption{Estimation error and latency of different latency-estimator variants.}
\label{table:estimator_variants}
\begin{tabular}{ccc}
\toprule
\textbf{Estimator Variant} & \textbf{Estimation Error} & \textbf{Latency} \\
\midrule
\systemname (additive)     & 2.01\%--2.81\% & 0.73 ms \\
Pairwise interaction model & 1.69\%--2.65\% & 0.95 ms \\
Multiplicative model       & 1.88\%--2.47\% & 1.04 ms \\
\bottomrule
\end{tabular}
\vspace{-.1in}
\end{table}

For latency, it is inversely proportional to the operating frequency of a processor~\cite{hennessy2011computer}: $t(f) = w_t \times \frac{1}{f}$. Applying similar regression leads to:
\begin{equation} \label{eqn:et}
    t_{i}(f_r) = (\gamma_r^0 + \sum\nolimits_{h=1}^{N_r} \gamma_r^h \psi_i^h) / f_r,
\end{equation}
where $\{\gamma_r^h\}$ are the regression parameters to be determined. The summation denotes an \textit{empirical equivalent-latency model}, not a serialized execution decomposition. Since GPU, CPU, and EMC activities may overlap during decoding, each term should be interpreted as the marginal sensitivity of end-to-end per-token latency to the processor frequency under hardware opacity.

To examine whether explicit concurrency modeling improves accuracy, we evaluated two concurrency-aware variants during the design. The first augments $\mathrm{LatE}_i$ with pairwise interaction terms ($\gamma_{r,r'}(\psi_i)/(f_r f_{r'})$ for each processor pair), and the second further adds a multiplicative term $\gamma_{r,r',r''}(\psi_i)/(f_r f_{r'} f_{r''})$. As shown in Table~\ref{table:estimator_variants}, the interaction-aware variants only marginally reduce the estimation error while increasing the estimator inference latency. Given that the governor repeatedly invokes the estimator during frequency search, we adopt the additive model for its better accuracy--overhead trade-off.

\begin{figure}[t]
    \begin{center}
        \includegraphics[width=.9\linewidth]{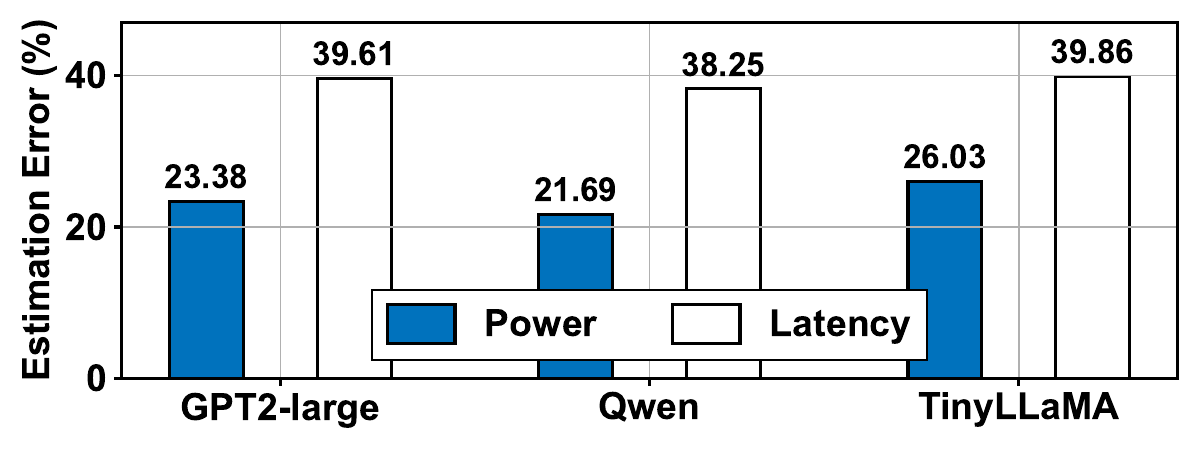}
    \end{center}
    \vspace{-.1in}
    \caption{Substantial errors of the estimated total power and latency performance when simply fitting all parameters for the three processors together.}
    \label{fig:e2e}
    \vspace{-.1in}

\end{figure}

\textbf{Parameter Determination.} Regression parameters for each processor can be derived through offline measurements of power consumption and latency across operating frequencies. These measurements enable direct fitting of $\{\lambda_r^h\}$ and $\{\gamma_r^h\}$ in Eqns.~(\ref{eqn:ep}) and (\ref{eqn:et}).

\textit{Hardware opacity.} Mobile edge devices expose only aggregated power and latency metrics across all three processors:
\begin{equation} \label{eqn:opacity}
    p_i(\{f_r\}) = \sum\nolimits_r p_i(f_r),~\text{and}~t_i(\{f_r\}) = \sum\nolimits_r t_i(f_r),
\end{equation}
which is theoretically sufficient for joint parameter fitting. However, we find this approach produces substantial estimation errors, for example, 21.69--26.03\% for power and 38.25--39.86\% for latency, as shown in Figure~\ref{fig:e2e}, which will lead to unsuitable and suboptimal frequency adjustments. 

Our analysis identifies the following reason. When all regression parameters are fitted together, the power and latency contributions of different processors are obscured during the co-execution of the \slm workload. The fitted parameters may not accurately reflect the actual behavior of each processor (maybe overfit to the training data), which naturally cannot produce accurate and reliable estimates in actual use after training.

\textbf{Decoupled Strategy.} To address hardware opacity, we observe that it is possible to decouple parameter fitting through controlled frequency isolation. Specifically, for each processor $r$, we can fix frequencies of the other two processors and vary $f_r$ to collect training data. This decoupling does not assume the GPU, CPU, and EMC execute independently or are physically isolated. Rather, it is a controlled parameter identification procedure under hardware opacity: by varying one processor frequency while fixing the others, we isolate the sensitivity of the aggregate measurement to that processor, which avoids the entangled and ill-conditioned fitting that causes the large errors. Taking power measurement as an example, it becomes:
\begin{equation}
    p_{i}(\{f_r\}) = (\lambda_r^0 + \sum\nolimits_{h=1}^{N_r} \lambda_r^h \psi_i^h) f_r^3 + p_i(others),
\end{equation}
where $p_i(others)$ refers to fixed baseline power due to other two processors. This isolates the parameters $\{\lambda_r^h\}$ of processor $r$, which we sequentially determine for all processors. Latency parameters $\{\gamma_r^h\}$ are similarly obtained.

\begin{figure}[t]
    \begin{center}
        \includegraphics[width=.94\linewidth]{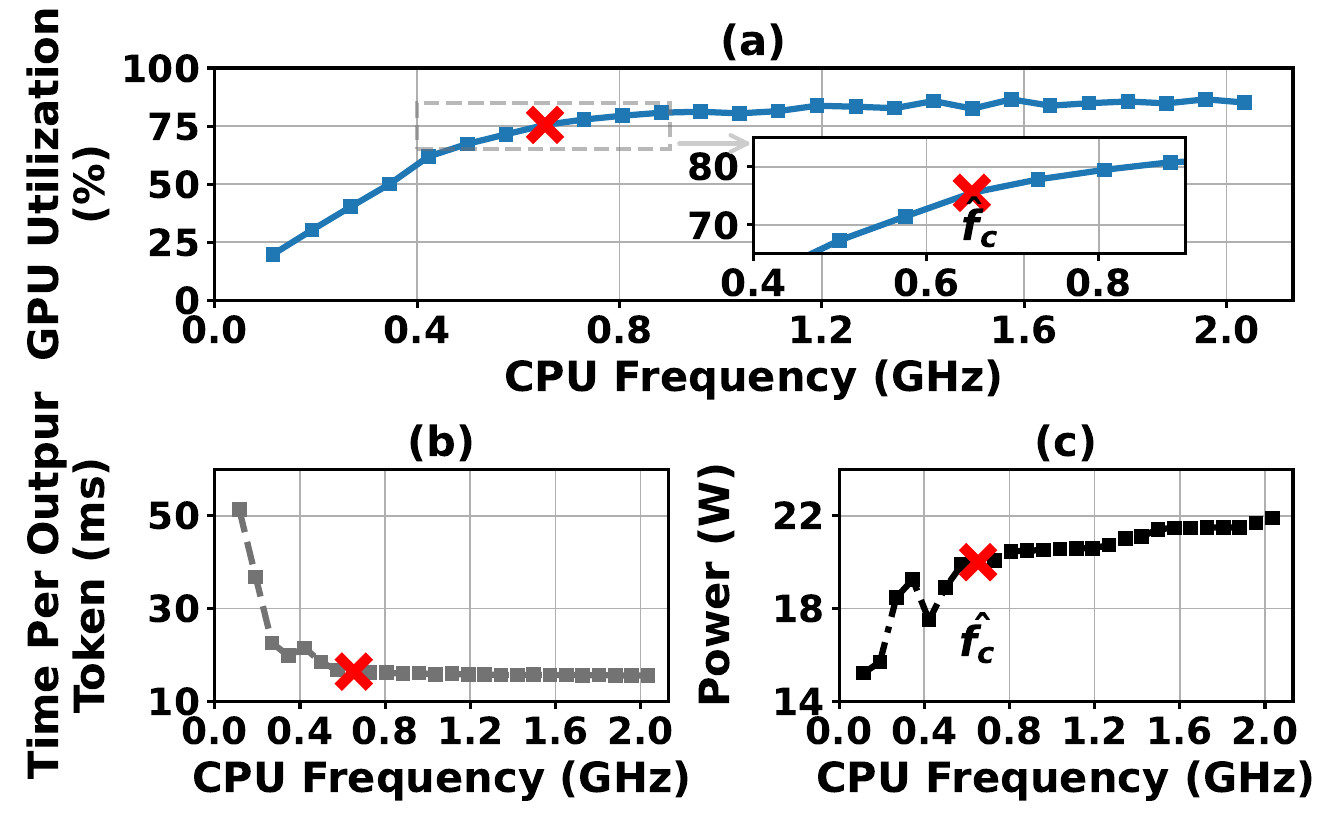}
    \end{center}
    \vspace{-.1in}
    \caption{(a) GPU utilization, (b) time per output token, and (c) power consumption when the CPU frequency is increased and the frequencies of GPU and EMC are fixed.}
    \label{fig:gpu_util}
    \vspace{-.1in}
\end{figure}

\textit{a) Latency estimator:} The final latency estimator, $\mathrm{LatE}_i$: $\{f_r\} \rightarrow t_{i}(\{f_r\})$ that estimates the total latency to generate each token $x_i$, combines per-processor models as follows:
\begin{eqnarray}
  \mathrm{LatE}_i: && \sum\nolimits_r (\gamma_r^0 + \sum\nolimits_{h=1}^{N_r} \gamma_r^h \psi_i^h) / f_r,
\end{eqnarray}
which achieves accurate estimation in the evaluation, \eg, the estimation error is only 2.01--2.81\% in Section~\ref{s:eval}.

\textit{b) Power estimator.} When similarly developing the power estimator, we observe significant estimation inaccuracies primarily caused by CPU power estimation at low operating frequencies. Specifically, the CPU plays a critical role in dispatching computation kernels to the GPU during \slm execution. At low CPU frequencies, delayed kernel dispatch causes stalled GPU waiting that decreases GPU utilization, as shown in Figure~\ref{fig:gpu_util}(a). Then, when increasing CPU frequency (\eg, from 0.1152 GHz to 0.6528 GHz), the total power consumption increases from two aspects: 
\begin{itemize}
    \item 1) inherent CPU power growth ($p \propto f^3$), and
    \item 2) increased GPU activation through enhanced kernel dispatch, evidenced by GPU utilization rising from 19.81\% to 75.51\%, as shown in Figure~\ref{fig:gpu_util}(a).
\end{itemize}
This GPU-CPU coupling effect is not captured by the individual processor models.

\systemname addresses this problem by exploiting the GPU-centric nature of \slm workloads. 
Our key observation is that there exists a critical CPU frequency threshold, denoted as $\hat{f}_c$. 
When the CPU frequency is below $\hat{f}_c$, GPU utilization remains low, leading to a strong GPU--CPU coupling effect and high latency. Once the CPU frequency exceeds $\hat{f}_c$, the GPU becomes sufficiently utilized, which substantially weakens this coupling effect. 
Further increasing the CPU frequency brings little latency benefit, while incurring additional power consumption, as illustrated in Figure~\ref{fig:gpu_util}(b) and (c). Therefore, fixing the CPU frequency at $\hat{f}_c$ is a principled design choice that balances latency and energy efficiency, rather than an arbitrary simplification.
Empirical measurements in Section~\ref{s:impl} further show that different \slms can share the same $\hat{f}_c$ (\eg, 0.6528 GHz in Table~\ref{table:slms}), making this threshold simple to configure in practice.

Since $\hat{f}_c$ resides within the low-frequency regime ($\hat{f}_c$ minimizes CPU power while sustaining GPU utilization), we keep the CPU frequency fixed at $\hat{f}_c$ and adjust only GPU and EMC frequencies during power governing. This yields our refined power estimator $\mathrm{PowE}_i$: $\{f_r\} \rightarrow p_{i}(\{f_r\})$ that estimates the total power consumption when generating each output token $x_i$ as:
\begin{eqnarray}
    \mathrm{PowE}_i:&&  \sum\nolimits_{r} (\lambda_r^0 + \sum\nolimits_{h=1}^{N_r} \lambda_r^h \psi_i^h) f_r^3,
\end{eqnarray}
where $r \in \{GPU, EMC\}$ in the current \systemname design. Direct modeling of GPU-CPU coupling remains an open problem for future work in this area.

\subsection{\systemname Governor}
\label{s:design:governor}

\begin{algorithm}[t]
\caption{\label{alg: governor} \systemname Governor}
\SetKwInOut{Input}{Input}
\begin{small}
\begin{spacing}{1.15}
\KwInput{Workload $\{\psi_i^h\}_{h=1}^M$; Deadline $\delta$; frequency sets $F_{GPU}$ and $F_{EMC}$; Adaptation factor $\alpha_k$;}
\SetAlgoVlined
\SetInd{0.05in}{0.1in}

$e_{min}\gets \inf;~\hat{f}_{g}\gets f_g^{max};~\hat{f}_{e}\gets f_e^{max}$\;

\For{$f_e$ \textbf{in} $F_{EMC}$}{
    $t_{min}\gets \alpha_k \times \mathrm{LatE}(\{\psi_i^h\}_{h=1}^M,f_g^{max},f_e)$\;
    $e_{lb}\gets \mathrm{PowE}(\{\psi_i^h\}_{h=1}^M,f_g^{min},f_e) \times t_{min}$\;
    \If{$t_{min}>\delta$ \textbf{or} $e_{lb}\ge e_{min}$}{
        \textbf{continue}\;
    }

    \For{$f_g$ \textbf{in} $F_{GPU}$}{
        $t\gets \alpha_k \times \mathrm{LatE}(\{\psi_i^h\}_{h=1}^M,f_g,f_e)$\;
        \If{$t>\delta$}{
            \textbf{continue}\;
        }

        $p\gets \mathrm{PowE}(\{\psi_i^h\}_{h=1}^M,f_g,f_e)$\;
        \If{$p\times t_{min}\ge e_{min}$}{
            \textbf{break}\;
        }

        \If{$p\times t<e_{min}$}{
            $e_{min}\gets p\times t;~\hat{f}_{g}\gets f_g;~\hat{f}_{e}\gets f_e$\;
        }
    }
}

$SetFreq(\hat{f}_{c},\hat{f}_{g},\hat{f}_{e})$\;

\end{spacing}
\end{small}
\end{algorithm}

Equipped with two estimators, $\mathrm{PowE}_i$ and $\mathrm{LatE}_i$, we now describe the \systemname governor's operation.

\textbf{Workflow.} The governor implements distinct power governing strategies for different inference phases:

\textit{a) Prefilling:} The CPU is fixed at $\hat{f}_{c}$ while the GPU and EMC operate at maximum frequencies to prioritize initialization speed.

\textit{b) Decoding:} The governor performs power governing once per decision window. Autoregressive decoding changes the workload after every generated token, so per-token scaling offers the finest granularity. However, scaling too frequently introduces non-negligible overhead. We therefore use a decision window of 50 tokens by default, which provides the best balance among responsiveness, energy efficiency, and latency QoS (Section~\ref{s:eval}). At the beginning of each window, before generating the next output token $x_i$, the governor
\begin{itemize}
    \item 1) obtains the metadata-parsed workload $\{\psi_i^h\}_{h=1}^M$,
    \item 2) evaluates all $\{f_{r}\}$ configurations (GPU/EMC frequencies) with CPU fixed at $\hat{f}_{c}$ using $\mathrm{PowE}_i$ for power estimates and $\mathrm{LatE}_i$ for latency estimates.
\end{itemize}

The governor then selects the frequencies $\{f_{r}\}$ that are optimal within the searched GPU--EMC space by solving the optimization problem in Eqn.~(\ref{eqn:opt}), as shown in Alg.~\ref{alg: governor}.

\textbf{Search acceleration.} To reduce the decision time, we precompute the latency and power frequency bases into lookup tables, and exploit the monotonicity within each EMC strip, where latency decreases and power increases as the GPU frequency rises. This enables three pruning rules: (1) discard an entire EMC strip if its minimum achievable latency already violates the deadline, or its lower-bound energy is worse than the current best; (2) skip the infeasible low-GPU-frequency prefix within each strip; and (3) terminate a strip scan once the current power combined with the strip-level minimum latency cannot improve the incumbent. Together, these rules reduce the average decision time from 1.73 ms to 0.81 ms across the evaluated \slms

\textbf{Infeasible deadlines.} When the requested deadline $\delta$ is so tight that no frequency candidates can satisfy it according to the latency estimator, the governor falls back to a best-effort policy: it raises all processors to their maximum frequencies. This maximizes the achievable latency QoS under the hardware limit.

\textbf{Adaptation.} While \slm execution typically dominates the workload of a mobile edge device, transient system disturbances or other concurrent tasks (\eg, image processing and DNNs) can impact estimator accuracy, subsequently affecting the efficiency of power governing. To ensure robustness, we introduce an window-level adaptation mechanism. For the $k$-th window $W_k$, the governor compares the accumulated estimated latency with the accumulated measured latency and computes an adaptation factor:
\begin{equation}
    \alpha_k =
    \frac{\sum_{i \in W_k} \hat{t}_i}
         {\sum_{i \in W_k} t_i},
\end{equation}
which serves as multiplicative correction to the latency estimates in the next window. This mechanism compensates for perturbations: increased observed latencies ($\alpha > 1$) prompt more aggressive frequency selections to maintain service quality, while reduced latencies ($\alpha < 1$) enable gradual power savings through more conservative selections.
\section{Implementation}
\label{s:impl}

\begin{table*}[t]
\centering
\resizebox{.93\linewidth}{!}{
\begin{tabular}{c c c c c c c c}
\hline
\hline
\textbf{Models} & \textbf{Params} & \textbf{\# of Layers} & \textbf{Hidden Size} & \textbf{MoE} & \textbf{Max Context Length} & $\mathbf{\hat{f}_c}$ \textbf{(GHz)} & \textbf{Release Date} \\
\hline
GPT2-small     & 117 M           & 12                    & 768                  & No                  & 1 k           &       0.6528       & 2019.12               \\
\hline
GPT2-medium    & 345 M           & 24                    & 1024                  & No                 & 1 k            &       0.6528                  & 2019.12               \\
\hline
GPT2-large     & 762 M            & 36                    & 1280                  & No                & 1 k            &      0.6528                   & 2019.12               \\
\hline 
GPT2-xl        & 1.5 B           & 48                    & 1600                  & No                 & 1 k            &        0.6528                 & 2019.12               \\
\hline
Qwen2.5-0.5B   & 0.5 B           & 24                    & 896                  & No                  & 32 k            &     0.6528                   & 2024.09               \\
\hline
TinyLLaMA      & 1.1 B            & 22                    & 2048                 & No                 & 2 k            &        0.6528                & 2024.03               \\
\hline
OLMoE-1B-7B    & 7 B           & 16                    & 2048                   & Yes                  & 4 k            &         0.6528               & 2024.09      \\ 
\hline
\end{tabular}
}
\caption{Details of the \slm models used in the evaluation. The GPT-2 series are used to evaluate the system performance with the same model structure but different parameters. Qwen2.5-0.5B and TinyLLaMA are two typical \slms. In addition, we further include an OLMoE-1B-7B based on the MoE architecture for comprehensive evaluation.}
\vspace{-.15in}
\label{table:slms}
\end{table*}

\parahead{Hardware}
We develop a prototype of \systemname and deploy it on NVIDIA Jetson AGX Orin. The frequencies of its GPU (NVIDIA Ampere), CPU (Arm Cortex-A78AE), and EMC are adjustable in the ranges of 0.3--1.3 GHz, 0.1--2.2 GHz, and 0.2--3.2 GHz, respectively. In addition, we deploy and test \systemname on Jetson Orin NX to further investigate its backward compatibility with lower-end devices. With the CPU fixed at $\hat{f}_c$, the runtime search reduces to the GPU--EMC space. On AGX Orin, this space contains 44 frequency combinations, while on Orin NX it contains 28 combinations. The search space differs across devices because each platform exposes a different set of available frequencies.

\parahead{Software}
We implement \systemname as a middleware layer between the kernel and applications. We develop the metadata parser using the XGBoost library~\cite{chen2016xgboost} and the power and latency estimators using sklearn.linear\_model. During the development, we also perform complementary tool development and code optimization, including developing a profiling toolkit on Linux Perf~\cite{de2010new} and NVIDIA CUDA Profiling Tools Interface (CUPTI)~\cite{cuptievent} to efficiently access hardware metadata, and excluding irrelevant input dimensions to the workload variation for the parser to facilitate XGBoost training. We use the built-in INA3221 to monitor device's power consumption. Since there are only four different frequency values for EMC, we use a different set of power estimation parameters for each EMC frequency. We develop the \systemname governor in C++ using XGBoost C API and expose both profiling module and governor to a Python API using the pybind11~\cite{pybind11} library for easy usage. For the inference framework of \slms, we employ llama.cpp~\cite{llamacpp} and its Python bindings llama-cpp-python~\cite{llamacpp-python}.

\parahead{Profiling cost} The profiling is a one-time offline calibration performed once per device--model pair. It does not require exhaustively profiling all CPU--GPU--EMC frequency combinations: since the CPU is fixed at the profiled threshold $\hat{f}_c$, we only profile the GPU--EMC combinations, and we sample only about 5\% of the context lengths (50 out of 1024). The detailed profiling time, storage, and energy costs are reported in Section~\ref{s:eval}.

\parahead{Energy measurement} To measure the energy consumption, we record the total token-generation latency $T$ of the inference process, and concurrently sample the device power with the built-in INA3221 power sensor at 200 Hz. The energy consumption is computed as $E = \bar{P} \cdot T$, where $\bar{P}$ is the average measured power during inference, and the energy per token is computed as $E / N$, where $N$ is the total number of generated tokens.

\parahead{SLM Models} 
To evaluate \systemname, we use seven representative \slms covering a wide range of model sizes, with the model parameters from 100+ M to 7 B, as shown in Table~\ref{table:slms}. Among them, the classic GPT2-large~\cite{radford2019language}, two more recent \slms Qwen2.5-0.5B~\cite{qwen2} and TinyLLaMA~\cite{zhang2024tinyllama}, and OLMoE-1B-7B~\cite{muennighoff2024olmoe} with MoE structure are used as the main models for evaluation. In addition, we also adopt other model variants for a more comprehensive evaluation, including different versions of GPT2, small, medium, large and extra large (xl), to explore the impact of model size, and reuse Qwen2.5-0.5B for long context lengths. We apply \texttt{q4\_0} quantization to these models for compatibility with the resource management schemes. We choose \texttt{q4\_0} because it is one of the most widely used and deployment-friendly quantization formats in llama.cpp-based edge inference, providing a good balance between memory footprint and inference quality. We emphasize that \systemname is orthogonal to the quantization scheme: \texttt{q4\_0} is only an example, and users may adopt other bit-widths (\eg, fp16) according to their accuracy and efficiency requirements. For each model, the $\hat{f}_c$ frequency of the CPU is selected empirically when the latency reduction per output token is less than 1.0\%. We find that the $\hat{f}_c$ frequency of all these \slms is the same, 0.6528 GHz, on both Orin and NX platforms, facilitating its configuration in practice.
\section{Evaluation}
\label{s:eval}

We compare the following DVFS methods in the evaluation:

\textbf{1) Rule-based (RULE)}: the latest built-in \dvfs governors on mobile edge devices. It employs the rule-based \texttt{nvhost\_podgov} for GPU, and \texttt{schedutil} for CPU, while EMC is held at a fixed frequency of 2.133 GHz~\cite{power}.

\textbf{2) GearDVFS (Gear)}: a state-of-the-art learning-based method~\cite{lin2023workload}, which designs a reinforcement learning-based agent to adjust processor frequencies. Since GearDVFS originally optimizes only utilization and temperature, we adapt it to \slm inference by adding a deadline compliance term to its reward. Let $L_j(\{f_r\})$ denote the token generation latency under frequencies $\{f_r\}$ and $\delta$ denote the deadline. The deadline compliance term is
\begin{equation}
    C(t) = 1 - \frac{|L_j(\{f_r\}) - \delta|}{\delta},
\end{equation}
which is maximized when the latency meets the deadline and decreases as the latency deviates from it. We retain the original utilization term $D_i(t)$ and thermal term $W_i(t)$ for each processor $i$, yielding the adapted reward
\begin{equation}
    R(t) = C(t) + \frac{1}{M}\sum\nolimits_{i=1}^{M}\left(D_i(t) + W_i(t)\right),
\end{equation}
and retrain the policy under this objective. For GearDVFS training, $\delta$ is set to 50 ms, corresponding to deadline rate of 20 tokens/s. After training, the policy is frozen and evaluated under different deadline rates. For a fair comparison, GearDVFS and \systemname are evaluated on the same hardware, workloads, deadline rates, and DVFS action space. GearDVFS observes the runtime signals used in its original design (utilization, frequencies, temperature, and measured token latency), while the estimator-specific features of \systemname (\eg, the matrix-operation scale) are not given to GearDVFS. Thus, each method uses the information assumed by its own design, while operating under an identical evaluation setup.

\begin{figure*}[t]
    \centering
    \begin{minipage}[t]{0.32\textwidth}
        \centering
        \includegraphics[width=\linewidth]{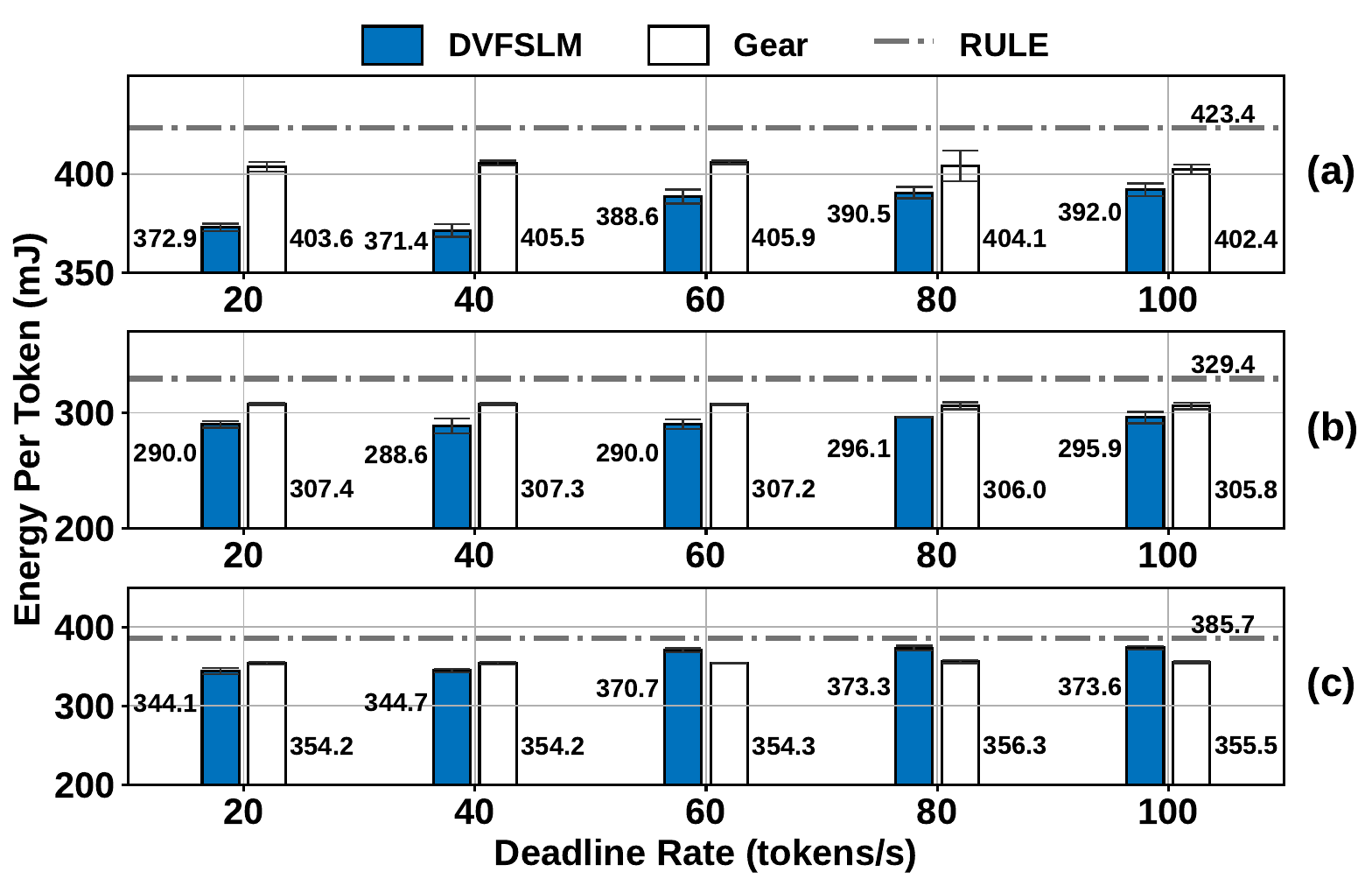}
        \vspace{-.2in}
        \caption{Energy per token of three methods at different deadline rates on (a) GPT2-large, (b) Qwen, (c) TinyLLaMA.}
        \label{fig:eval:ept}
    \end{minipage}
    \hfill
    \begin{minipage}[t]{0.32\textwidth}
        \centering
        \includegraphics[width=\linewidth]{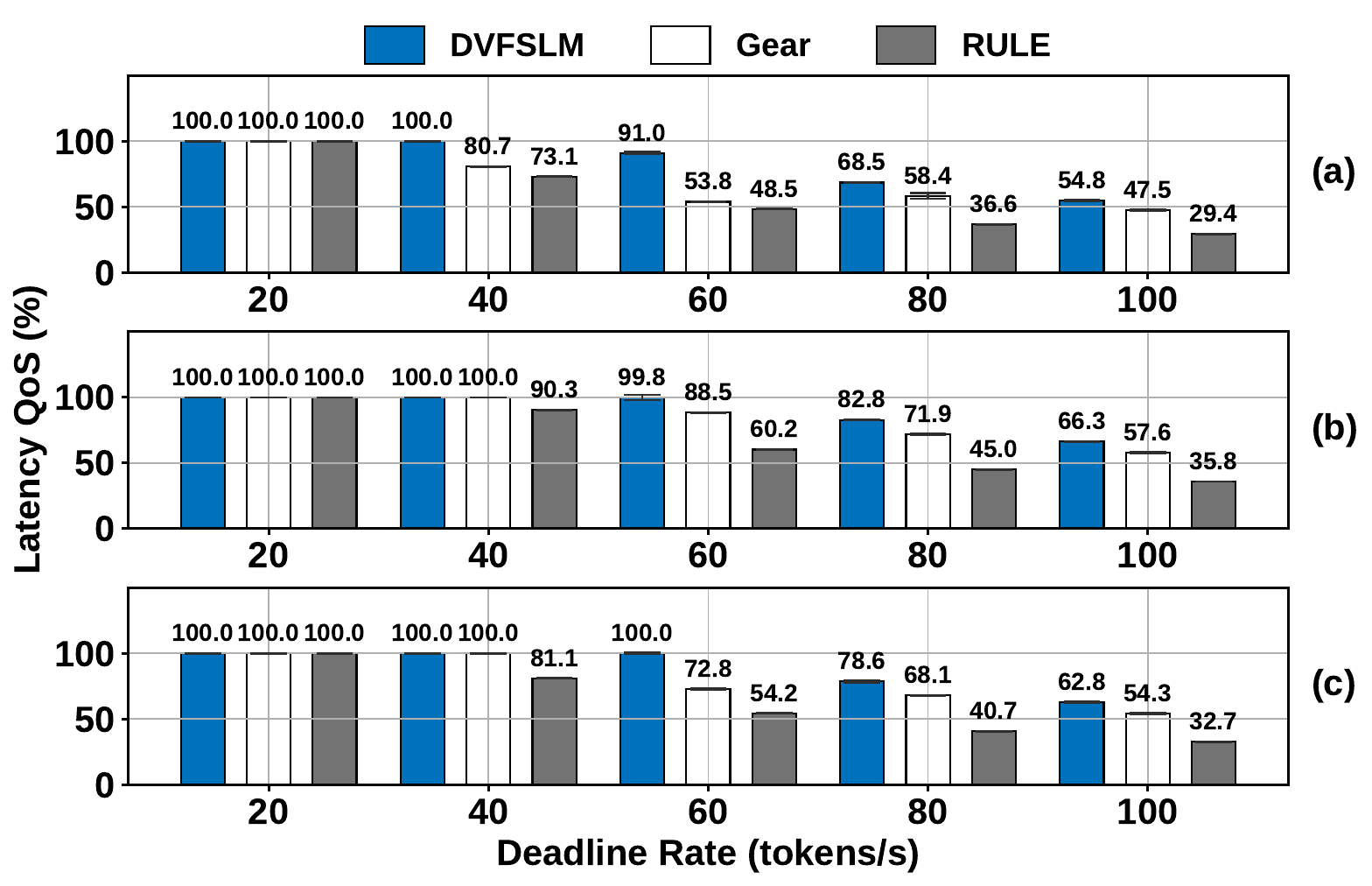}
        \vspace{-.2in}
        \caption{Latency QoS of each method at different deadline rates on (a) GPT2-large, (b) Qwen, and (c) TinyLLaMA.}
        \label{fig:eval:qos}
    \end{minipage}
    \hfill
    \begin{minipage}[t]{0.32\textwidth}
        \centering
        \includegraphics[height=0.68\linewidth]{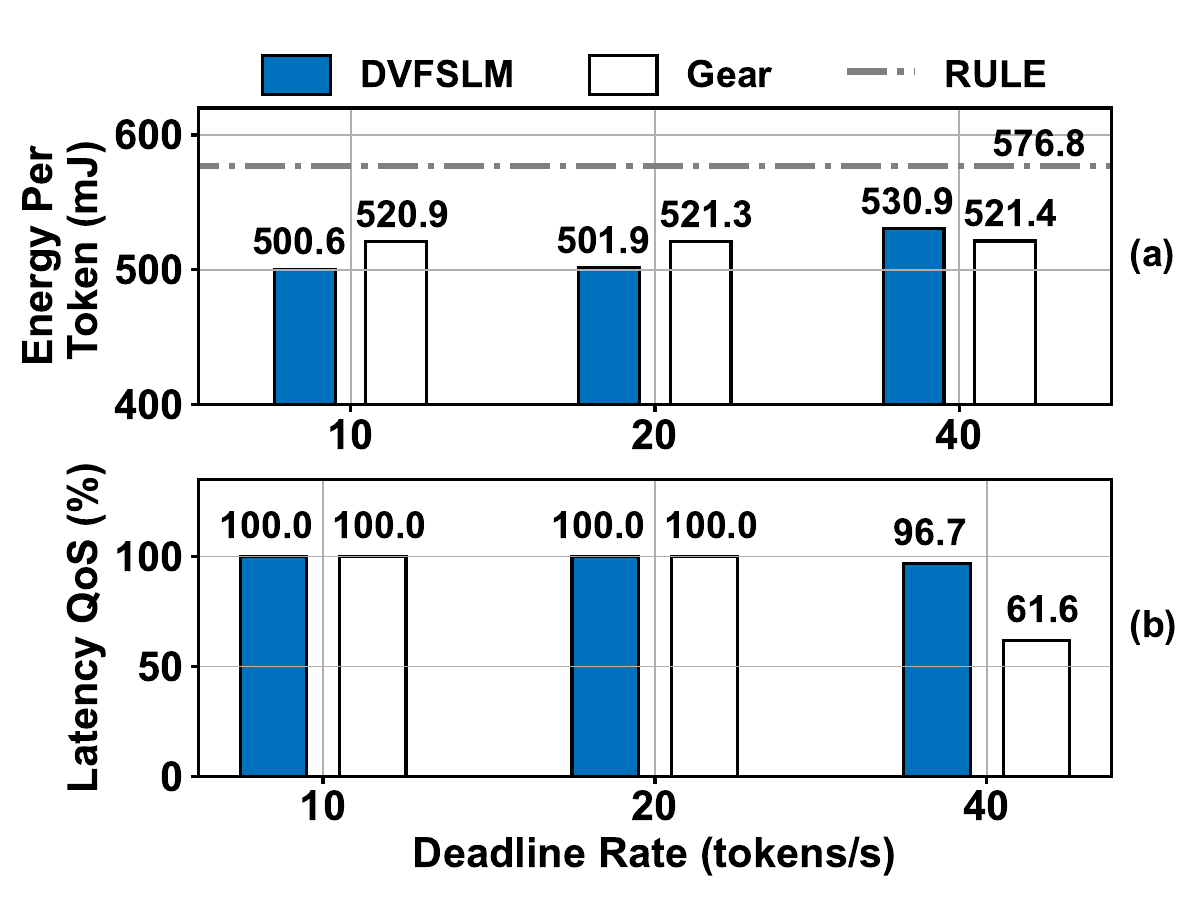}
        \vspace{-.03in}
        \caption{(a) Energy per token and (b) latency QoS of each method at different deadline rates on OLMoE.}
        \label{fig:eval:moe}
    \end{minipage}
    \vspace{-.14in}
\end{figure*}

\textbf{3) \systemname}: the method proposed in this paper. 

In the evaluation, the following two metrics are used to quantify the performance of each method:

\textbf{1) Energy per Token (EPT):} this metric quantifies the energy efficiency of \slms and is calculated as the total energy consumption during decoding divided by the total number of tokens generated. The lower this metric is, the more energy-efficient the \slm is.

\textbf{2) Quality of Service (QoS)}: this metric quantifies the service quality of \slm inference, calculated as QoS $ = \frac{r_{a}}{r_{d}}$, where $r_{a}$ and $r_{d}$ are the \textit{achieved rate} and the \textit{deadline rate} (\ie, the minimum rate required by the user or application), respectively. If $r_{a} > r_{d}$, the latency QoS cap is 100\%.

\subsection{Overall Performance}
\label{s:eval:overall}

We compare the performance of each method when running GPT2-large, Qwen, TinyLLaMA, and OLMoE. In this experiment, we run each \slm to generate 1k (1024) output tokens, and will evaluate the performance when generating more tokens in the micro-benchmark. Figure~\ref{fig:eval:ept} and Figure~\ref{fig:eval:qos} show the energy and latency performance of each method, respectively. We repeat each experiment 5 times and report the mean together with the 95\% confidence interval (represented with error bars), computed as $\bar{x} \pm t_{\alpha/2,\,n-1}\cdot s/\sqrt{n}$, where $\bar{x}$ and $s$ are the sample mean and standard deviation, $n=5$, and $t_{\alpha/2,\,n-1}$ is the critical value of the $t$-distribution at the 95\% confidence level. The resulting confidence intervals are narrow, indicating that the performance of the proposed method is stable across runs. The standard deviation of \systemname's energy per token remains under 2.52, while that of its latency QoS is below 1.56. The target deadline rates include both feasible and infeasible rates that exceed the peak token generation rate of the device.

\parahead{Energy per token} As shown in Figure~\ref{fig:eval:ept}, RULE is inefficient when applied to \slms, resulting in relatively high energy consumption of 329.4 to 423.4 mJ. Since RULE cannot change the generation rate of output tokens, we plot its per-token energy consumption as a line in each sub-graph. Gear has the potential to run at different deadline rates and is more energy efficient than RULE. However, its design fails to capture the unique and complex workload characteristics of \slms well, resulting in limited energy efficiency and latency performance (to be shown). Due to the effective design, \systemname can improve both aspects at the same time. From Figure~\ref{fig:eval:ept}, we can see that \systemname outperforms RULE and Gear by up to 12.4\% and 8.4\% respectively. 

As shown in Figure~\ref{fig:eval:ept}(c), for the deadline rate of 60 to 100 tokens/s on TinyLLaMA, Gear's per-token energy consumption appears lower than \systemname. By digging deeper into the experiment logs, we find that this result actually shows that Gear has limitations on \slms. For such a high deadline rate, \dvfs should use higher frequencies to provide more computing power. However, Gear fails to do so and results in serious deadline violations.

\parahead{Latency QoS} We then study the latency performance of each setting involved in Figure~\ref{fig:eval:ept}. Figure~\ref{fig:eval:qos} shows that when the deadline is low (\eg, 20 tokens/s), all methods achieve rates above the deadline rate. When the deadline becomes tighter (\eg, 40 tokens/s), Gear and RULE start to miss deadlines. When the deadline rate is even higher (\eg, over 60 tokens/s), Gear chooses conservative (but inaccurate) frequencies, which do not waste energy but result in a higher deadline miss rate. In contrast, \systemname can adapt to different deadline requirements and outperform RULE and Gear by up to 93.12\% and 69.14\% respectively.

\parahead{Infeasible deadlines} Figure~\ref{fig:eval:ept} and Figure~\ref{fig:eval:qos} also show results for infeasible deadline rates (\eg, 80 and 100 tokens/s), which exceed the peak token rate the device can sustain. As described in Section~\ref{s:design:governor}, \systemname will fall back to a best-effort policy that minimizes the predicted token latency. Even under these infeasible deadlines, \systemname still attains higher latency QoS than Gear while keeping the energy per token competitive, as its frequency selection is guided by the latency estimator and reacts to the changing workload, whereas Gear fails to push the rate to the device limit.

\parahead{Performance of MoE} We further evaluate the performance of each method when running MoE-based \slm OLMoE. Since OLMoE has much more parameters than the \slms used in the previous experiments, we reduce the deadline rate to 10, 20, and 40 in this experiment. In Figure~\ref{fig:eval:moe}, we observe similar conclusions. RULE achieves the lowest energy efficiency when running OLMoE. Gear improves the energy consumption per token to 520.9--521.4 mJ, but at the cost of a low latency QoS of 61.6\% at the highest deadline rate (40 tokens/s). Unlike Gear, \systemname can ensure both high energy efficiency and latency QoS performance.

\subsection{Efficacy of Technical Modules}

\begin{figure*}[t]
    \centering
    \begin{minipage}[t]{0.32\textwidth}
        \centering
        \includegraphics[width=\linewidth]{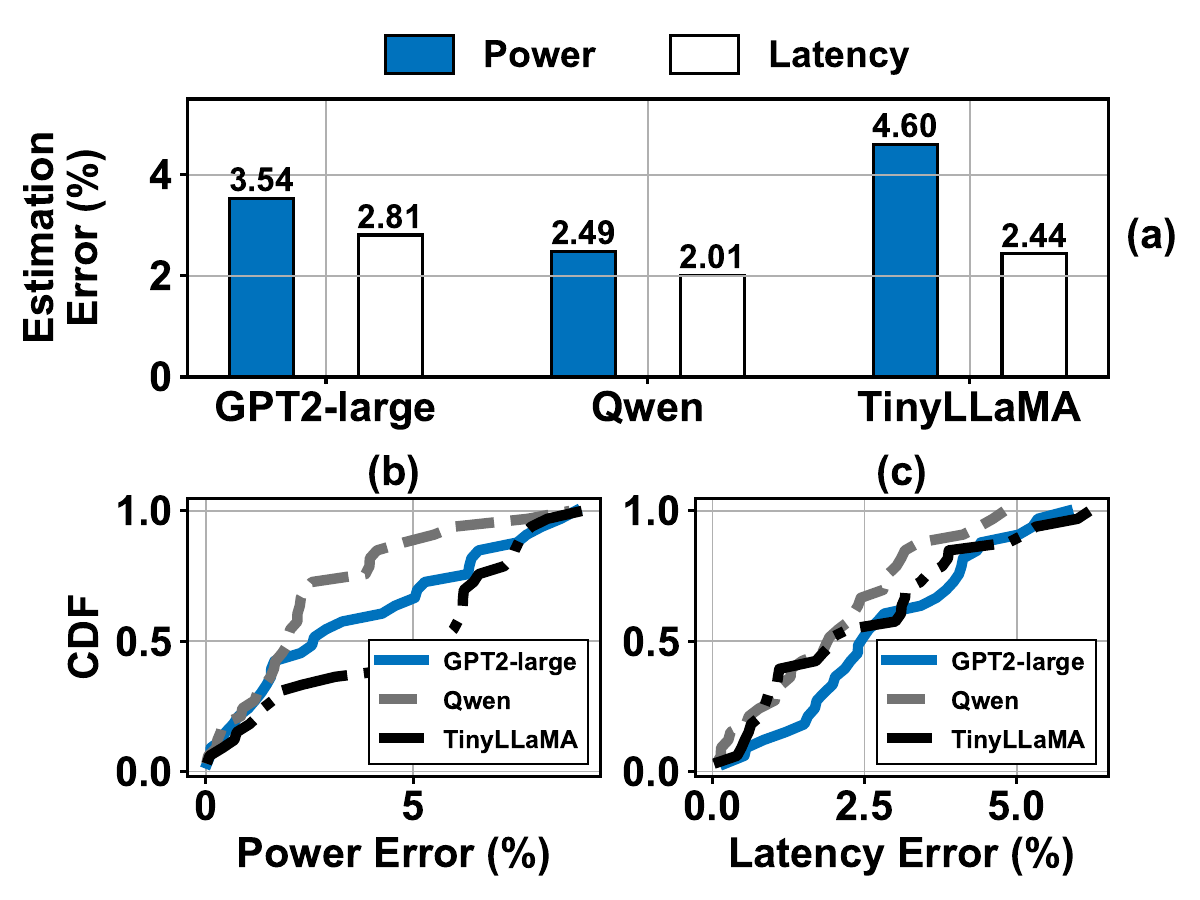}
        \vspace{-.2in}
        \captionof{figure}{(a) Estimation error of power and latency estimators on three \slms. CDF of (b) the power and (c) latency estimations for each \slm.}
        \label{fig:eval:analysis}
    \end{minipage}
    \hfill
    \begin{minipage}[t]{0.32\textwidth}
        \centering
        \includegraphics[width=\linewidth]{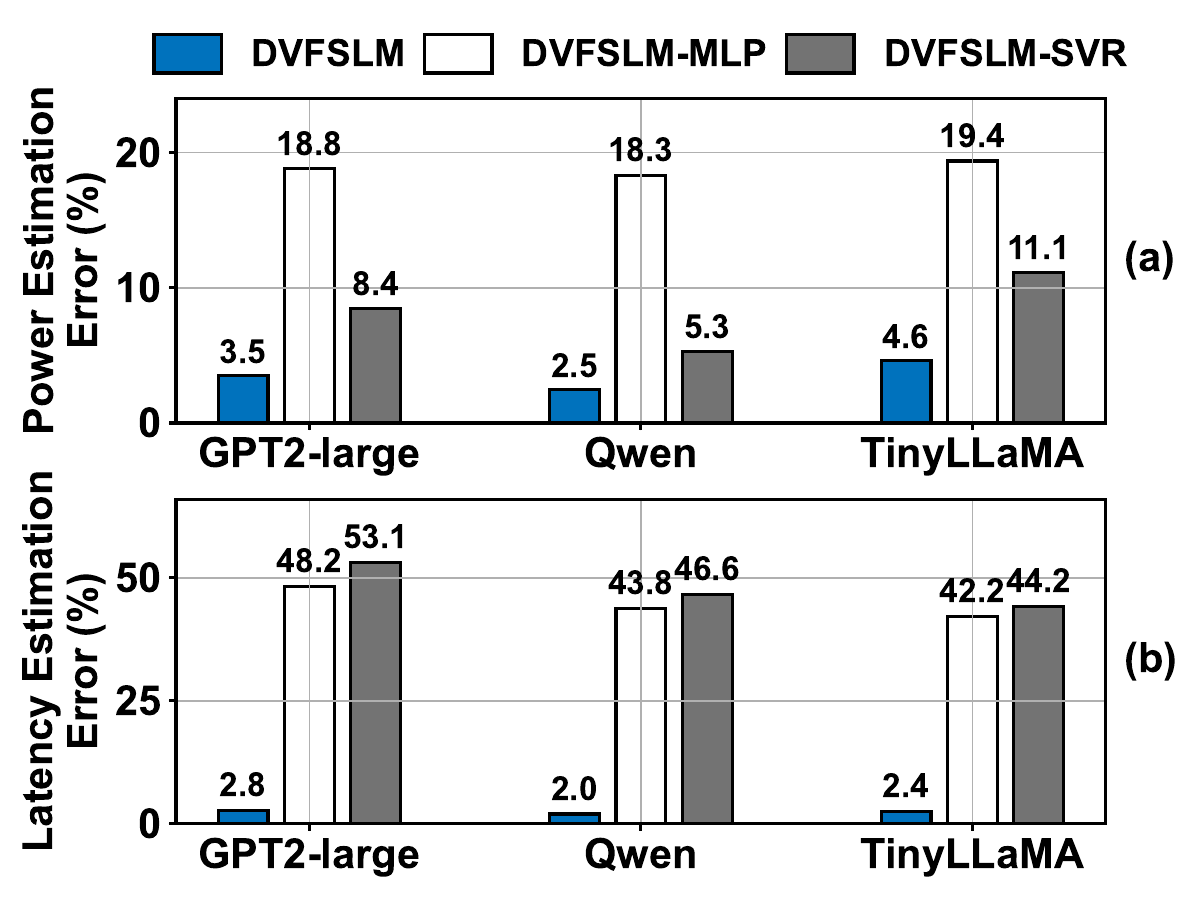}
        \vspace{-.2in}
        \captionof{figure}{Ablation study of (a) power and (b) latency estimation errors using different approaches to design estimators.}
        \label{fig:eval:estimation}
    \end{minipage}
    \hfill
    \begin{minipage}[t]{0.32\textwidth}
        \centering
        \includegraphics[width=\linewidth]{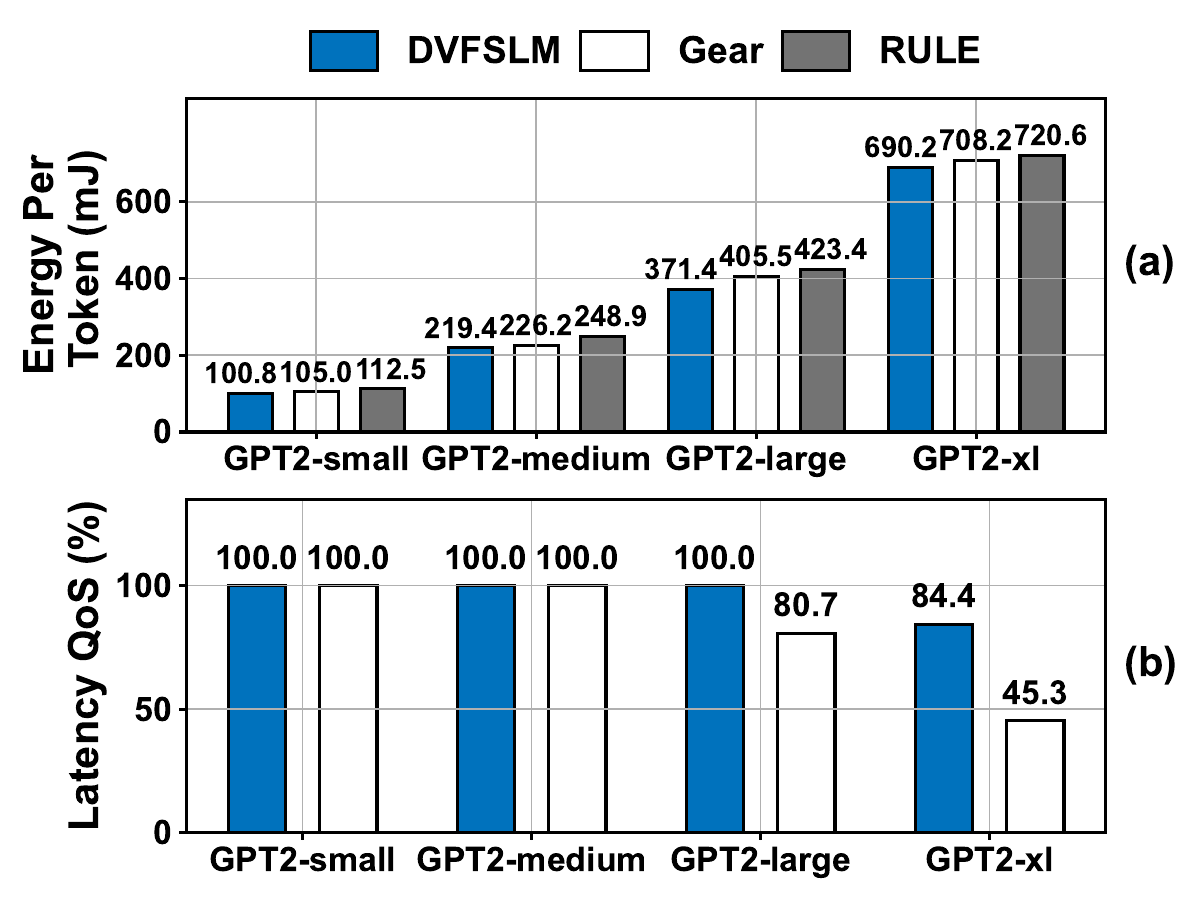}
        \vspace{-.2in}
        \captionof{figure}{Impact of model size. (a) Energy per token and (b) latency QoS on different variants of the GPT2 model.}
        \label{fig:eval:model_size}
    \end{minipage}
    \vspace{-.14in}
\end{figure*}

\begin{figure*}[t]
    \centering
    \begin{minipage}[t]{0.31\textwidth}
        \centering
        \includegraphics[width=\linewidth]{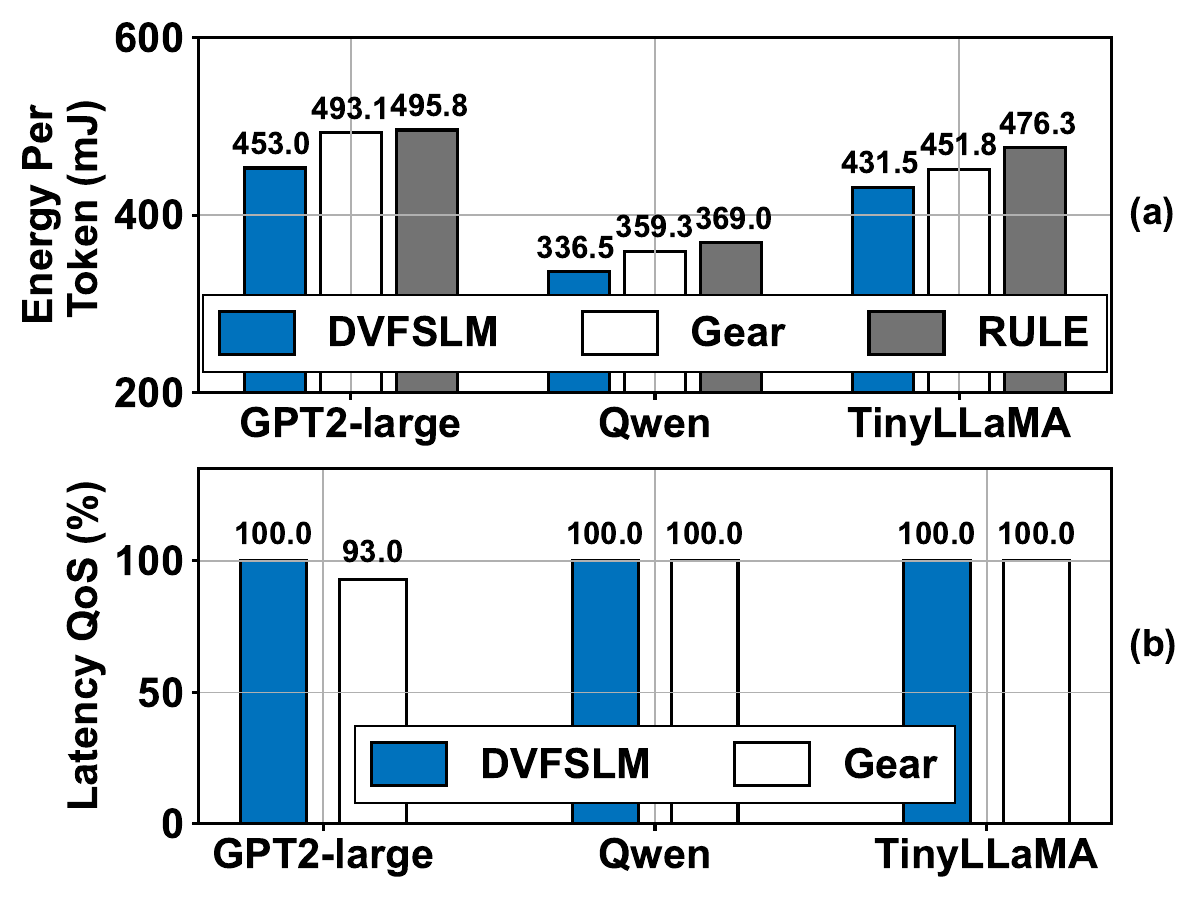}
        \vspace{-.2in}
        \captionof{figure}{(a) Energy per token and (b) QoS for each method when running GPT2-large on a lower-end device NX.}
        \label{fig:eval:device}
    \end{minipage}
    \hfill
    \begin{minipage}[t]{0.31\textwidth}
        \centering
        \includegraphics[width=\linewidth]{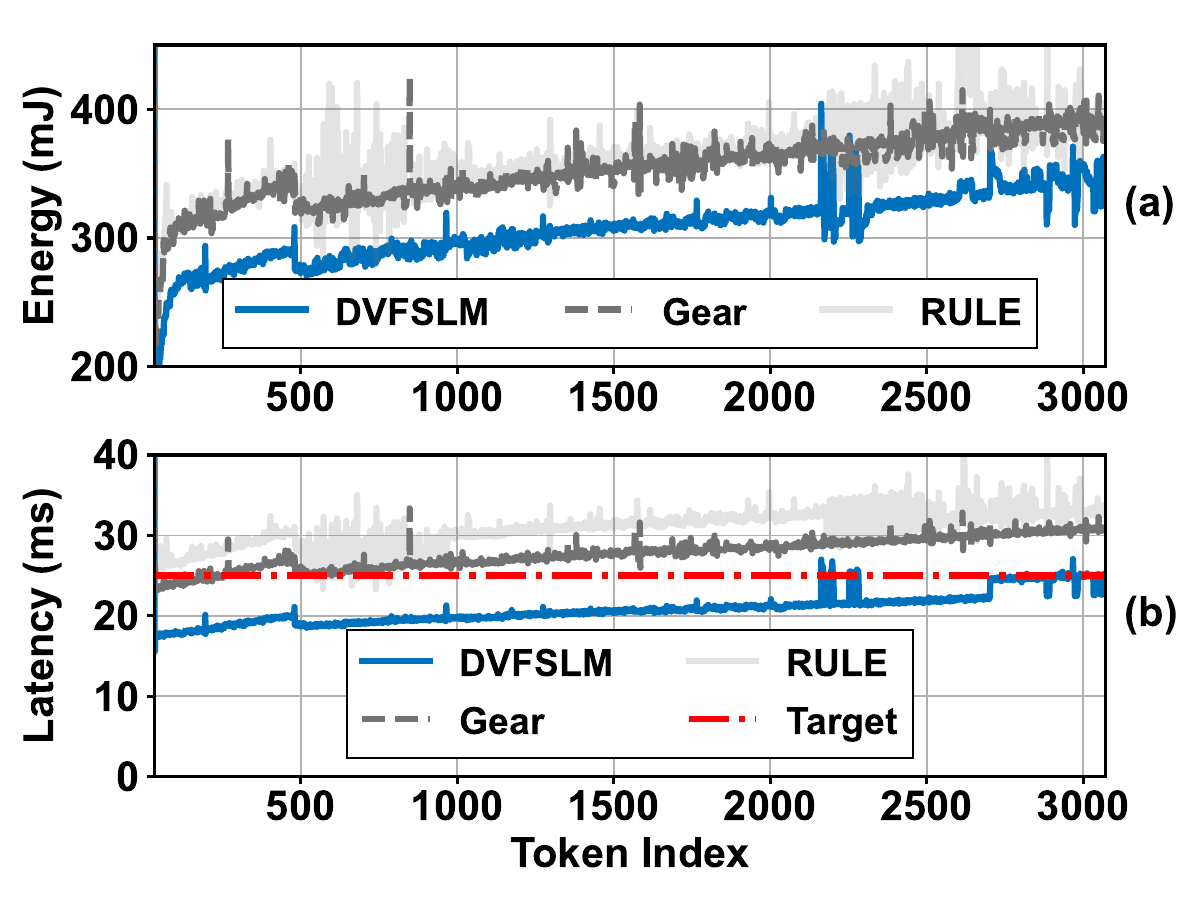}
        \vspace{-.2in}
        \captionof{figure}{(a) Energy per token and (b) latency per token for each method when running Qwen with long context length.}
        \label{fig:eval:long context} 
    \end{minipage}
    \hfill
    \begin{minipage}[t]{0.34\textwidth}
        \centering
        \includegraphics[width=\linewidth]{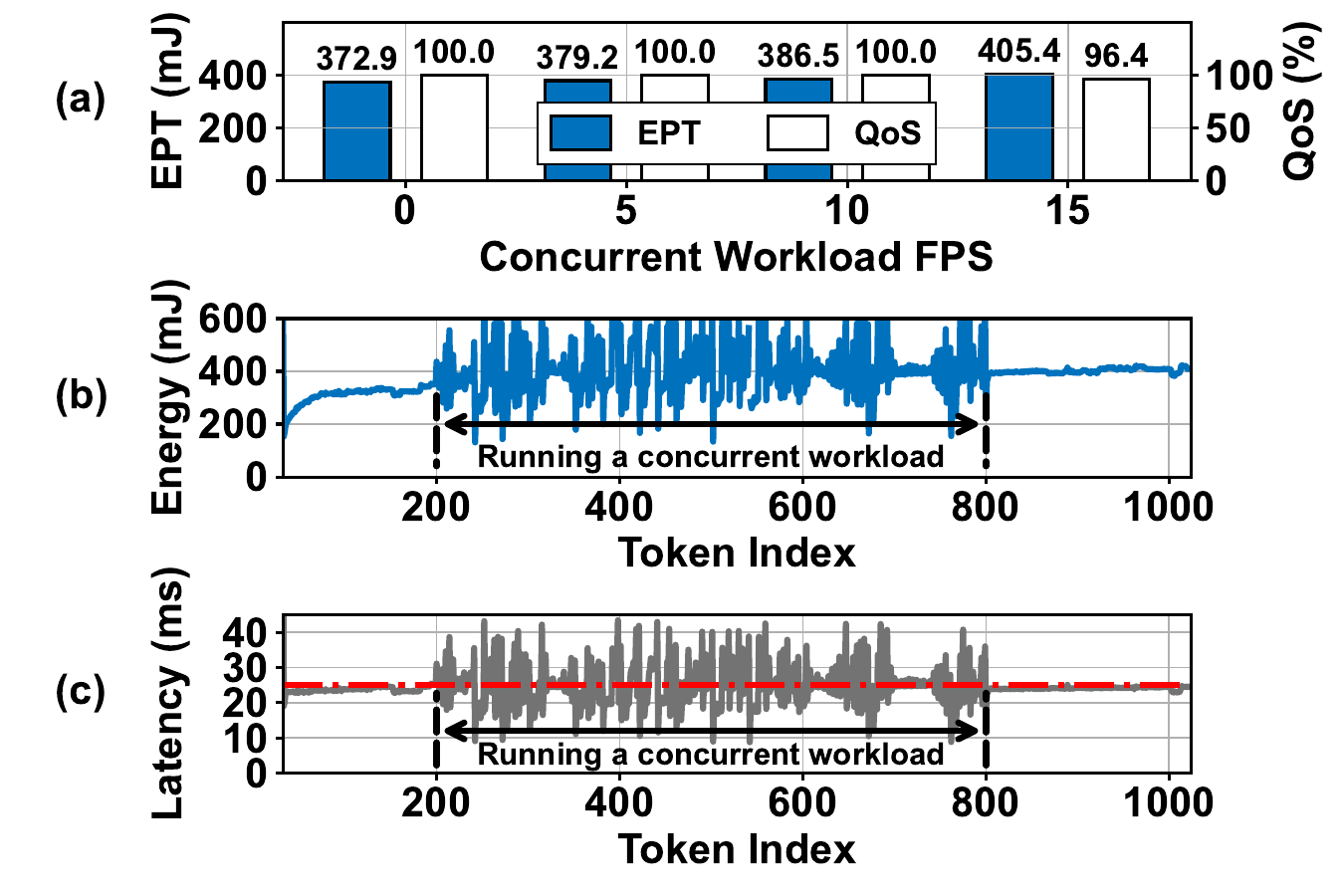}
        \vspace{-.2in}
        \captionof{figure}{(a) Energy per token and QoS when MobileNet runs alongside GPT2-large at 5/10/15 FPS. (b) Energy and (c) latency when MobileNet runs at 10 FPS.}
        \label{fig:eval:concurrent}
    \end{minipage}
    \vspace{-.14in}
\end{figure*}

To better understand the performance of \systemname achieved above, we dig into the performance of the key technical modules proposed in \systemname in this subsection.

\parahead{Performance of estimators} The power governing of \systemname relies on the accuracy of power and latency estimation at different frequency combinations. Figure~\ref{fig:eval:analysis}(a) depicts the power and latency estimation errors of different \slms used in the previous section. We can see that \systemname achieves low error rates, ranging from 2.49\% to 4.60\% for power and 2.01\% to 2.81\% for latency on average, showing their effectiveness in driving \systemname's frequency selections. Figure~\ref{fig:eval:analysis}(b) and (c) show the detailed error distribution of power and latency for each \slm.

\parahead{Ablation study} When designing the power and latency estimators of \systemname, we first explore their underlying principles, which provide the opportunity to realize the design using a simple regression model. In this ablation study, we develop two additional versions of \systemname to verify the necessity of our design by replacing our estimator design with end-to-end learning using two common models: ``\systemname-MLP'' using Multilayer Perceptron (MLP) and ``\systemname-SVR'' using Support Vector Regression (SVR).

As shown in Figure~\ref{fig:eval:estimation}, the original \systemname estimator consistently outperforms the designs in \systemname-MLP and \systemname-SVR, especially achieving more significant improvements in latency estimation. Overall, the original estimator reduces the latency estimation error by 94.1\% to 95.7\% and the power estimation error by 52.8\% to 86.3\%. These experimental results demonstrate the effectiveness of our estimator design in \systemname.

\subsection{Micro-Benchmarks}
\label{s:eval:micro}

To comprehensively evaluate how the performance of \systemname is affected by system configurations, we conduct the following micro-benchmark experiments.

\parahead{Different model sizes} To examine the impact of different model sizes, we evaluate each method on different variants of GPT2, including GPT2-small (117 M), GPT2-medium (345 M), GPT2-large (762 M), and GPT2-xl (1.5 B), with gradually increasing numbers of model parameters. This experiment uses a medium deadline rate of 40 tokens/s. Figure~\ref{fig:eval:model_size}(a) shows that for small models like GPT2-small, \systemname outperforms the other two methods by 4.0--10.4\% and also achieves meaningful improvements for large models, such as 18.0--52.0 mJ per token that yield 2.5--12.3\% improvements for GPT2-large and GPT2-xl. Across all models, \systemname improves the energy efficiency per token by 9.7\% and 4.5\% over RULE and Gear, respectively. Meanwhile, \systemname consistently achieves higher latency QoS relative to Gear, with the average latency of all models improving by 27.6\%, as illustrated in Figure~\ref{fig:eval:model_size}(b).

\parahead{Different devices} In this experiment, we further deploy \systemname on Jetson Orin NX. Since its computing power is lower than AGX Orin, we evaluate each method with a deadline rate of 20 tokens/s. From Figure~\ref{fig:eval:device}(a), we can see that the energy consumption per token of \systemname ranges from 336.5 to 453.0 mJ, outperforming RULE and Gear by 8.6--9.4\% and 4.5--8.1\%, respectively. Figure~\ref{fig:eval:device}(b) further shows that \systemname meets the latency QoS requirements of all three models on the Orin NX device.

\begin{figure*}[t]
    \centering
    \begin{minipage}[t]{0.33\textwidth}
        \centering
        \includegraphics[width=\linewidth]{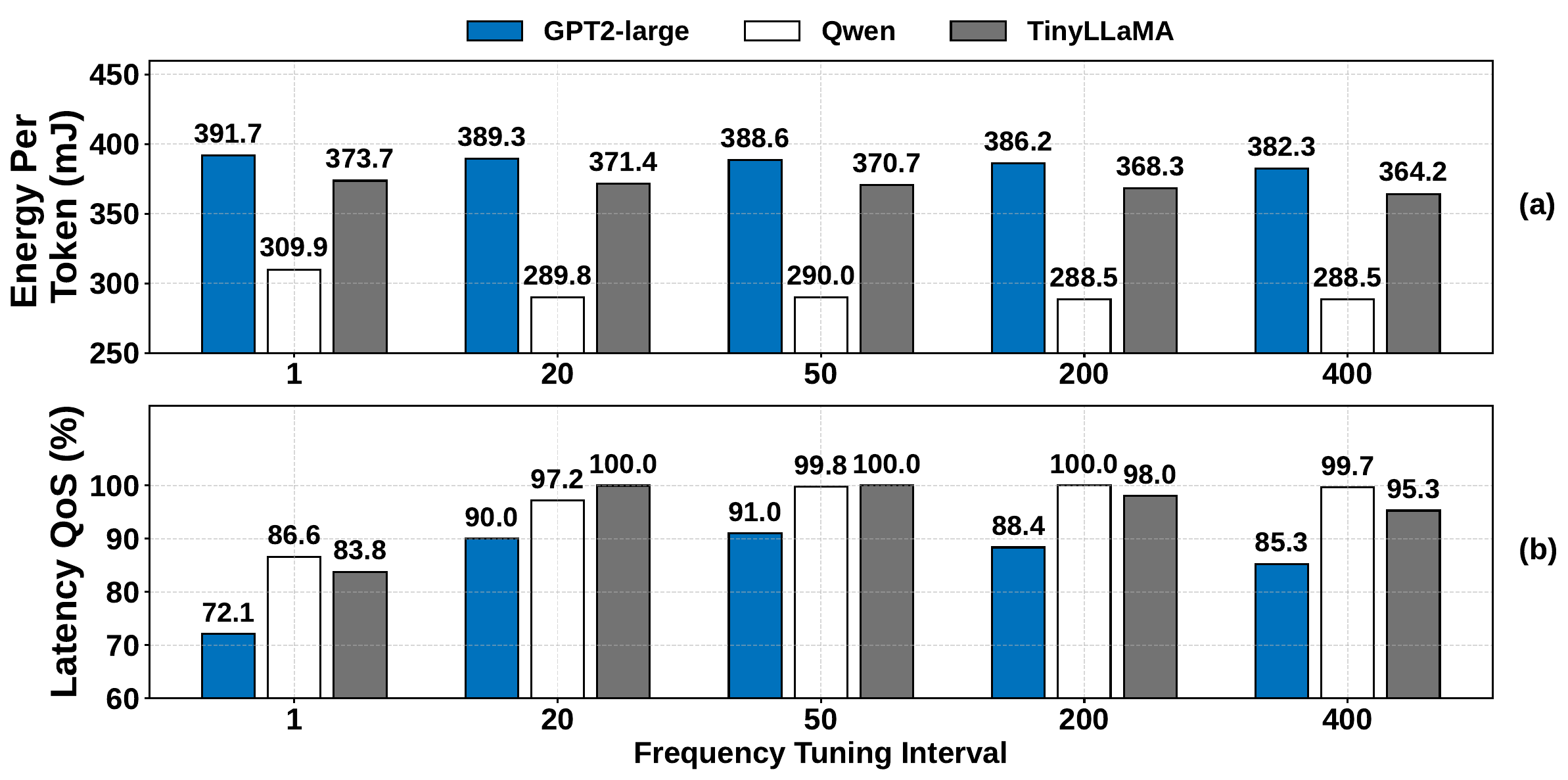}
        \vspace{-.2in}
        \captionof{figure}{(a) Energy per token and (b) latency QoS under different frequency-scaling intervals across different models.}
        \label{fig:eval:interval}
    \end{minipage}
    \hfill
    \begin{minipage}[t]{0.33\textwidth}
        \centering
        \includegraphics[width=\linewidth]{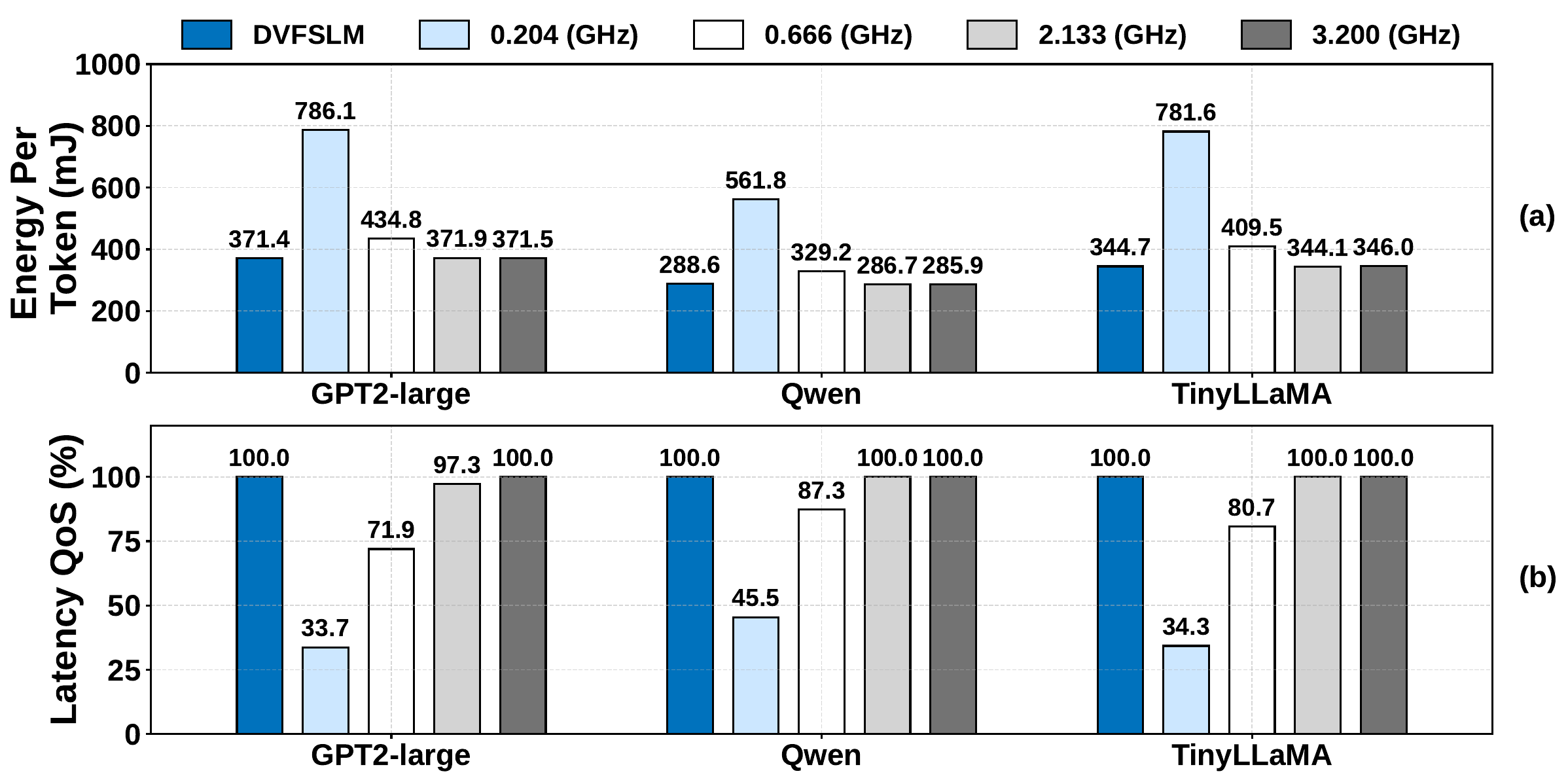}
        \vspace{-.2in}
        \captionof{figure}{(a) Energy per token and (b) latency QoS under varying EMC frequencies across different models.}
        \label{fig:eval:emc}
    \end{minipage}
    \hfill
    \begin{minipage}[t]{0.3\textwidth}
        \centering
        \includegraphics[width=\linewidth]{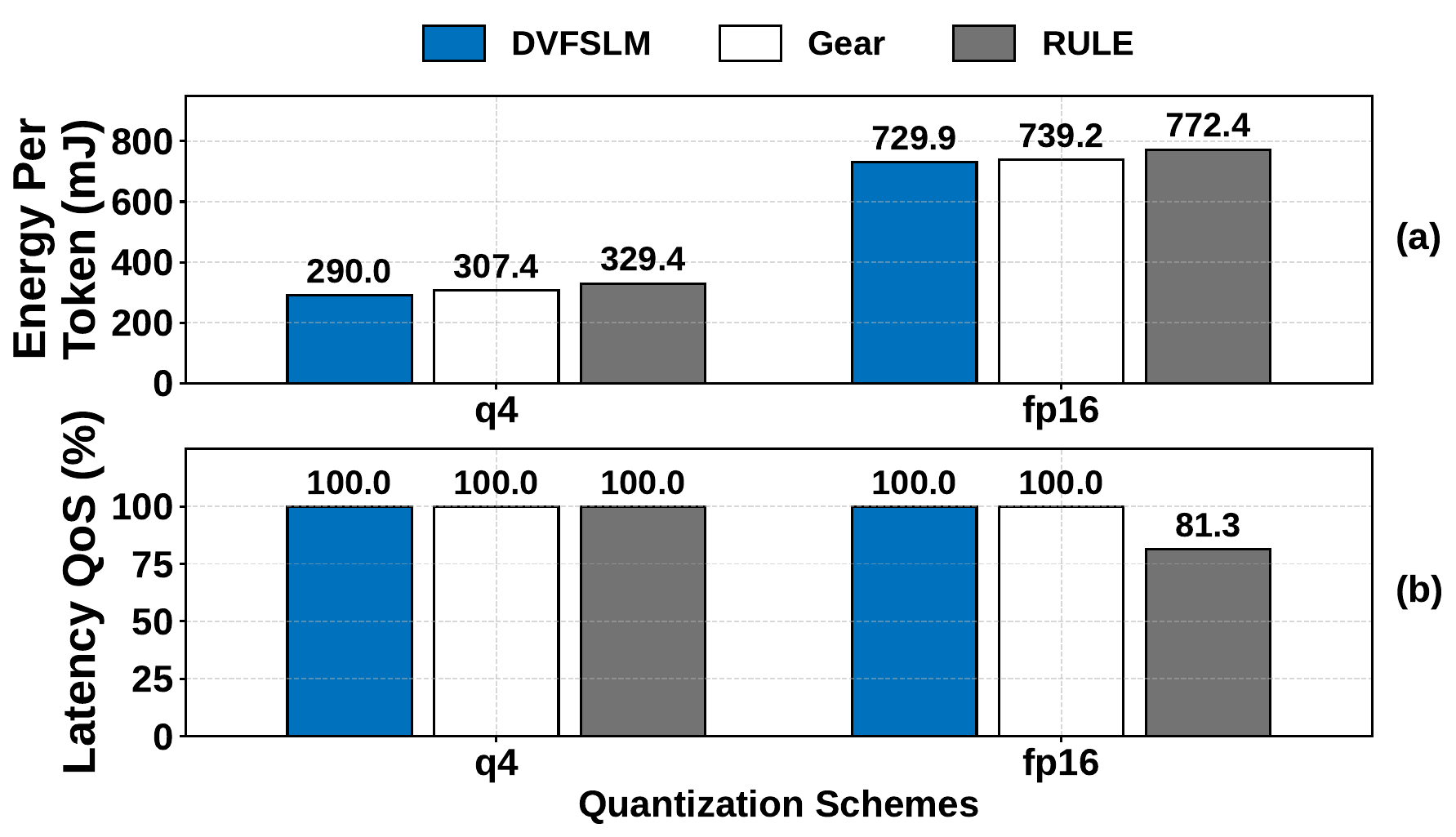}
        \vspace{-.2in}
        \captionof{figure}{(a) Energy per token and (b) latency QoS for different quantization schemes on Qwen.}
        \label{fig:eval:quant}
    \end{minipage}
    \vspace{-.14in}
\end{figure*}

\parahead{Long context length} In \slm-based applications, long outputs can be generated, and we evaluate the performance of \systemname with longer context lengths. In Figure~\ref{fig:eval:long context}, we run the decoding stage of Qwen with up to 3k tokens and a deadline rate of 40 tokens/s. As mentioned earlier, autoregression gradually increases the workload, and the energy per token of each method in Figure~\ref{fig:eval:long context}(a) also increases accordingly, where \systemname can improve the average energy consumption of RULE and Gear by 14.6\% and 13.0\%, respectively. On the other hand, \systemname adheres to the deadline rate well, and the actual latency of generating each output token is relatively stable and is lower than the latency corresponding to the deadline rate (25 ms = $\frac{1 ~\text{second}}{40~\text{tokens/s}}$) most of the time, as illustrated in Figure~\ref{fig:eval:long context}(b).

\begin{figure*}[t]
    \centering
    \begin{minipage}[t]{0.33\textwidth}
        \centering
        \includegraphics[width=\linewidth]{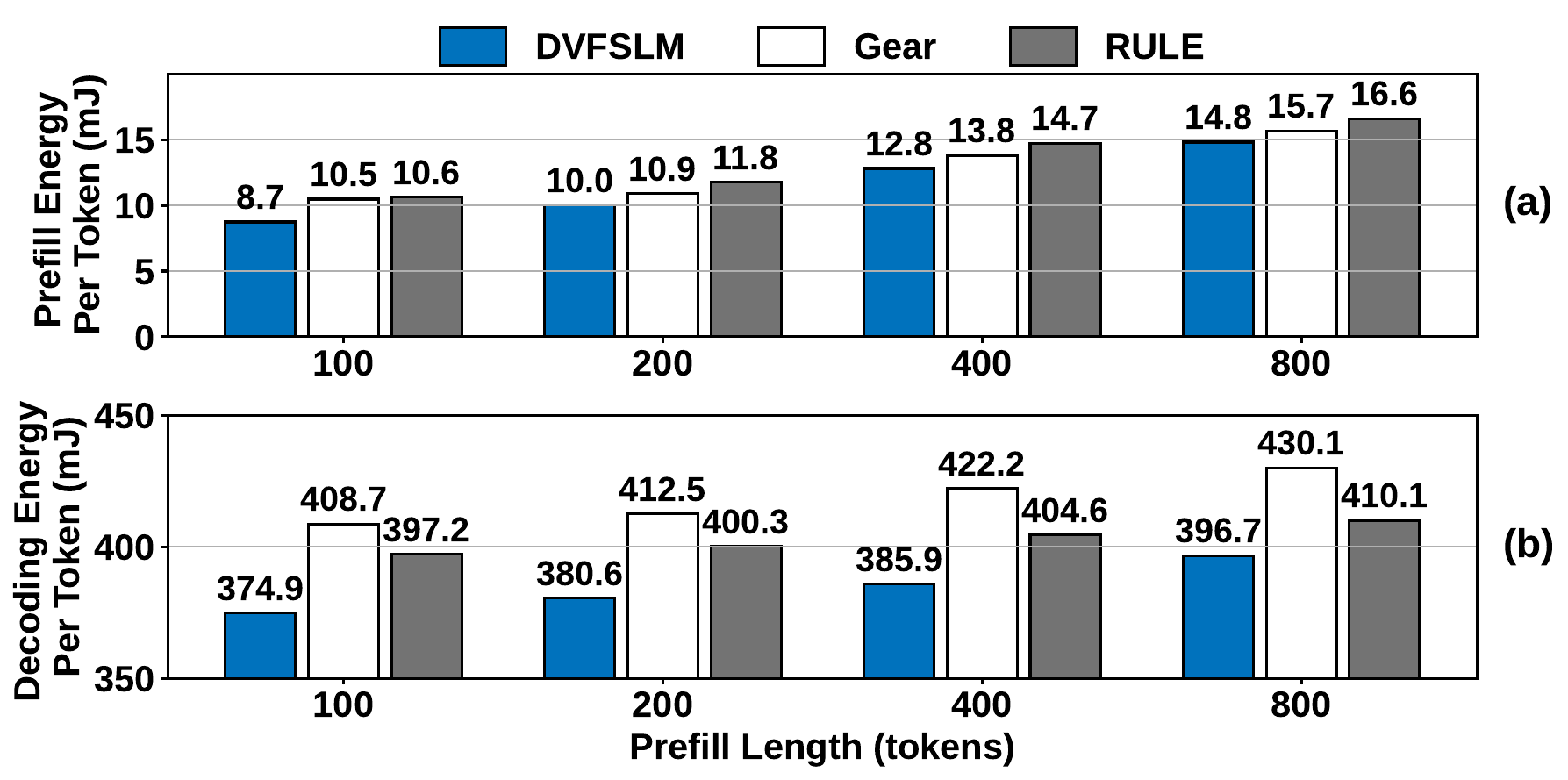}
        \vspace{-.2in}
        \captionof{figure}{Energy per token of each method on (a) prefill and (b) decoding at different prefill lengths.}
        \label{fig:eval:prefill_energy}
    \end{minipage}
    \hfill
    \begin{minipage}[t]{0.33\textwidth}
        \centering
        \includegraphics[width=\linewidth]{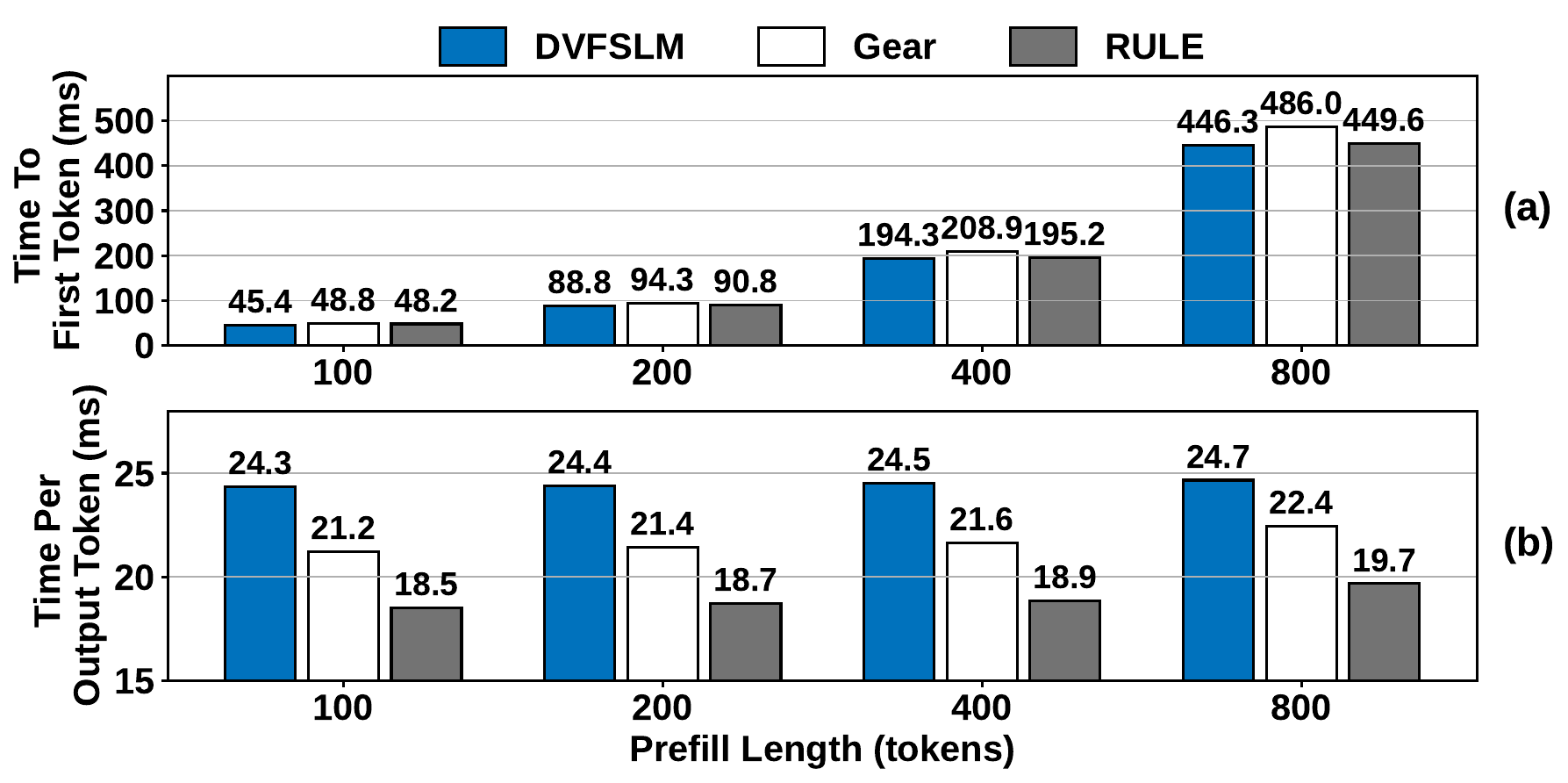}
        \vspace{-.2in}
        \captionof{figure}{Latency of each method on (a) prefill and (b) decoding at different prefill lengths.}
        \label{fig:eval:prefill_time}
    \end{minipage}
    \hfill
    \begin{minipage}[t]{0.29\textwidth}
        \centering
        \includegraphics[width=\linewidth]{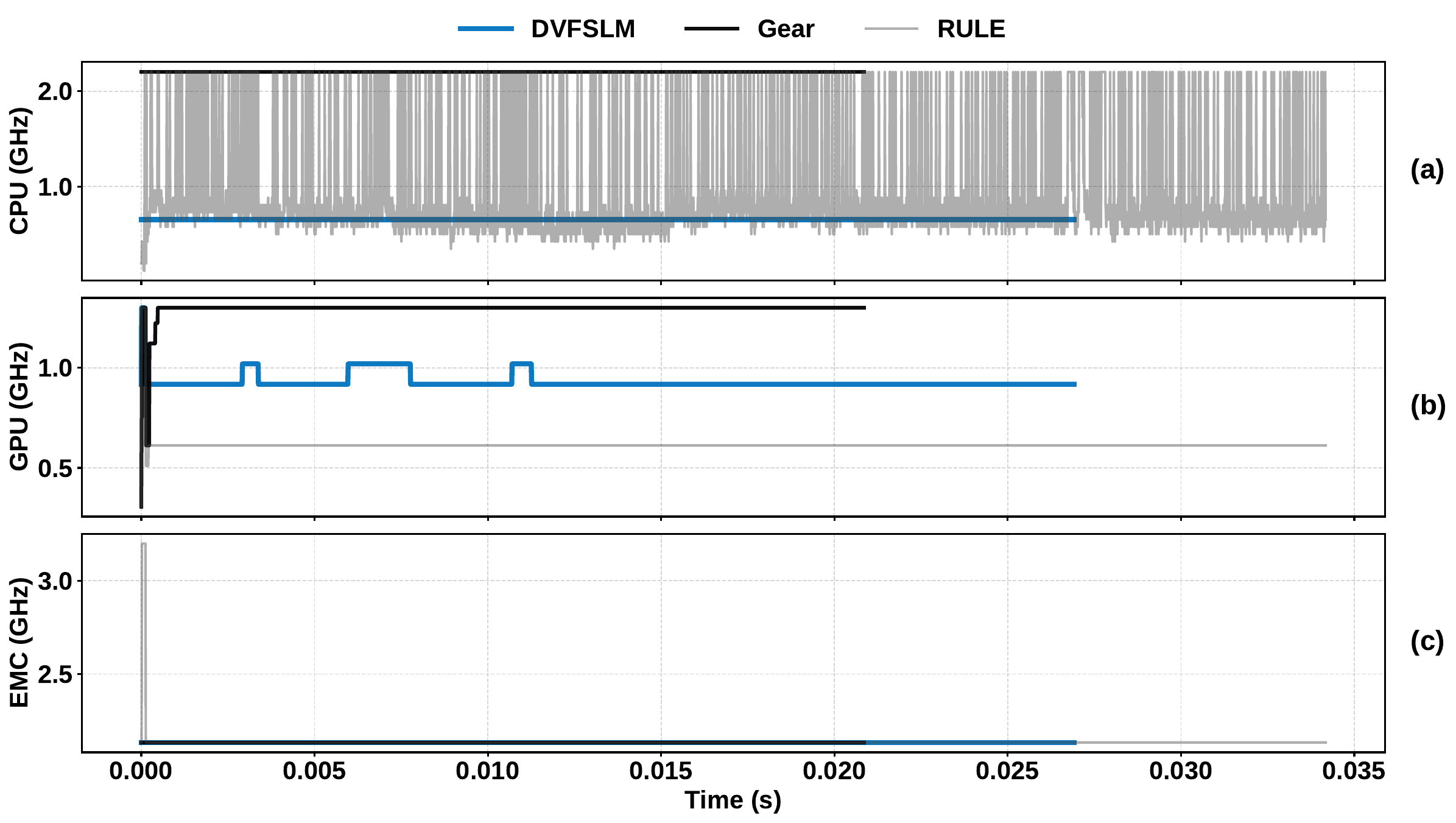}
        \vspace{-.2in}
        \captionof{figure}{Time-series visualization of frequency decisions when running GPT2-large.}
        \label{fig:eval:freqtrace}
    \end{minipage}
    \vspace{-.14in}
\end{figure*}

\parahead{Concurrent workload} On mobile edge devices, other tasks may run alongside \slms, and these additional workloads may affect power and latency estimates of \systemname, thus affecting its power governing. To evaluate the effectiveness of our adaptive design for this issue (Section~\ref{s:design:governor}), we run MobileNet concurrently alongside GPT2-large (deadline rate: 40 tokens/s), and change the intensity of this additional workload by increasing its frame rate from 10 to 15 frames per second (FPS). We also show the performance of \systemname without running MobileNet as a reference (0 FPS).

As shown in Figure~\ref{fig:eval:concurrent}(a), the energy consumption increases slightly because a higher frequency is required to process the additional workload while meeting the latency QoS requirements. Due to the effective adaptation, \systemname can maintain 100\% high latency QoS at 5--10 FPS and 96.4\% high latency QoS at 15 FPS. In Figure~\ref{fig:eval:concurrent}(b) and (c), we plot the runtime energy consumption and latency performance per token when MobileNet runs concurrently at 10 FPS. The energy consumption and latency fluctuate due to the dynamic changes of the concurrent workload. However, every time the generation of an output token slows down, the adaptive mechanism reacts quickly to speed up the generation of the next token so that the overall generation rate meets the deadline. In the future, we plan to explore applying filtering techniques to smooth out such fluctuations.

\parahead{Frequency-scaling interval} \systemname performs frequency scaling once per decision window. To justify the default 50-token window (Section~\ref{s:design:governor}), we vary the scaling interval and set the deadline rate to 60 tokens/s for each model. As shown in Figure~\ref{fig:eval:interval}, scaling every single token incurs higher energy per token and lower latency QoS, because of the overhead of overly frequent frequency adjustments. Using a 50-token interval instead achieves latency QoS of 91.0\%, 99.8\%, and 100.0\% for GPT2-large, Qwen, and TinyLLaMA, respectively, while also reducing the energy per token across all three models. Increasing the interval further makes the governor less responsive to workload changes and degrades latency QoS. Therefore, the 50-token interval provides the best trade-off among responsiveness, energy efficiency, and latency QoS.

\parahead{EMC frequency sensitivity} Since \slm decoding requires frequent accesses to model weights and KV-cache data, the EMC frequency is particularly important. To study its impact, we manually set the EMC frequency to different levels while keeping the CPU and GPU frequencies selected by \systemname, and evaluate GPT2-large, Qwen, and TinyLLaMA at a deadline of 40 tokens/s. As shown in Figure~\ref{fig:eval:emc}, the EMC frequency significantly affects both energy per token and latency QoS. When the EMC frequency is too low (\eg, 0.204 GHz), all models suffer severe QoS degradation due to insufficient memory bandwidth, and the energy per token also rises because the prolonged execution time offsets the lower EMC power. Increasing the EMC frequency improves QoS, but an excessively high EMC frequency does not always reduce energy. Since \systemname models the EMC frequency in its power and latency estimators (Section~\ref{s:design}), it can select EMC frequencies that satisfy the latency QoS while minimizing energy. Due to estimation errors, dynamic environment, and the need to maintain QoS, DVFSLM may choose a slightly more conservative EMC setting. Thus, a fixed EMC frequency can achieve slightly lower energy per token than DVFSLM in some cases.

\parahead{Quantization sensitivity} We choose \texttt{q4\_0} as the default quantization scheme, but \systemname is orthogonal to the quantization choice. To verify this, we compare \texttt{q4\_0} with the fp16 setting on Qwen at a deadline rate of 20 tokens/s. As shown in Figure~\ref{fig:eval:quant}, \systemname achieves consistent gains in energy per token and latency QoS under both quantization schemes. Experiments on the WikiText-2 dataset show that the fp16 model achieves a perplexity of 13.91, while the \texttt{q4\_0} model achieves a perplexity of 15.66, corresponding to a moderate degradation of 12.58\%. In practical deployment, users can choose different quantization schemes according to their accuracy and efficiency requirements, and \systemname can adjust its frequency decisions accordingly to meet the latency QoS while minimizing energy.

\begin{figure*}[t]
    \centering
    \begin{minipage}[t]{0.33\textwidth}
        \centering
        \includegraphics[width=\linewidth]{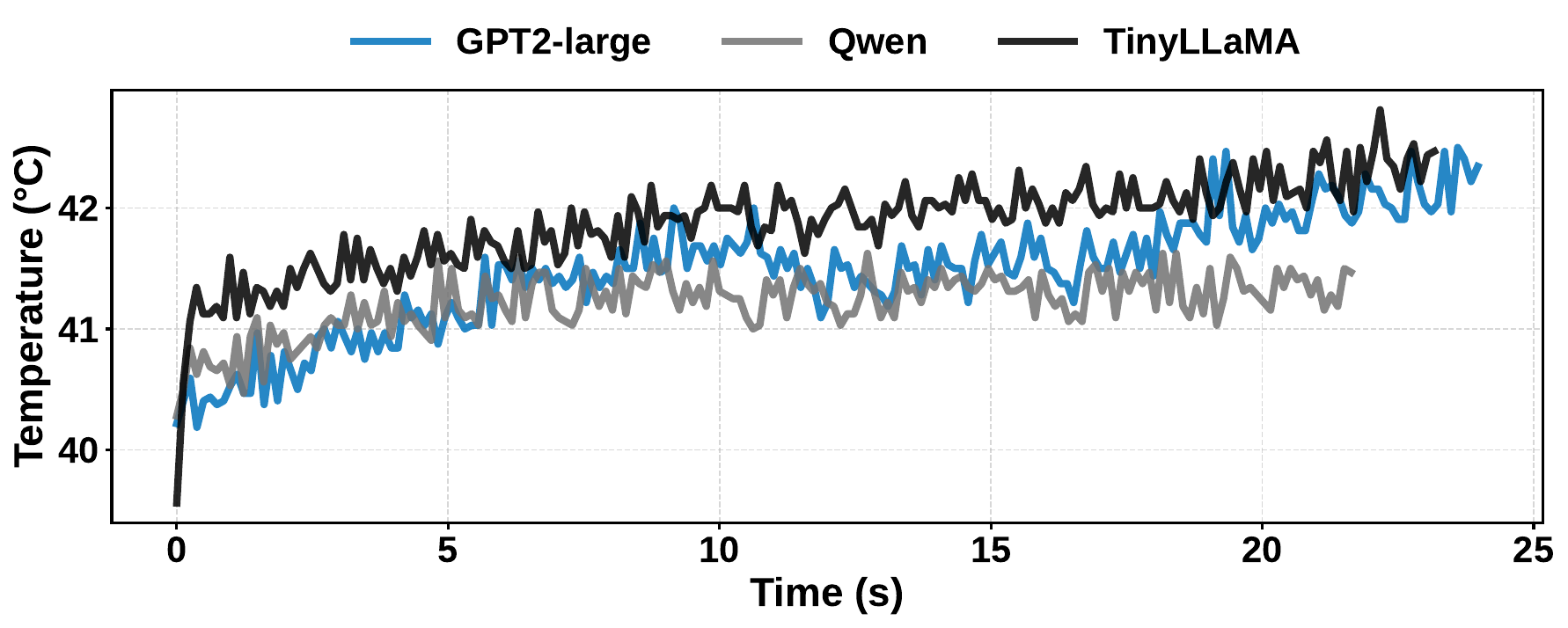}
        \vspace{-.2in}
        \captionof{figure}{Device temperature under an ambient temperature of 25\,$^{\circ}$C when running different \slms with \systemname.}
        \label{fig:eval:temp}
    \end{minipage}
    \hfill
    \begin{minipage}[t]{0.30\textwidth}
        \centering
        \includegraphics[width=\linewidth]{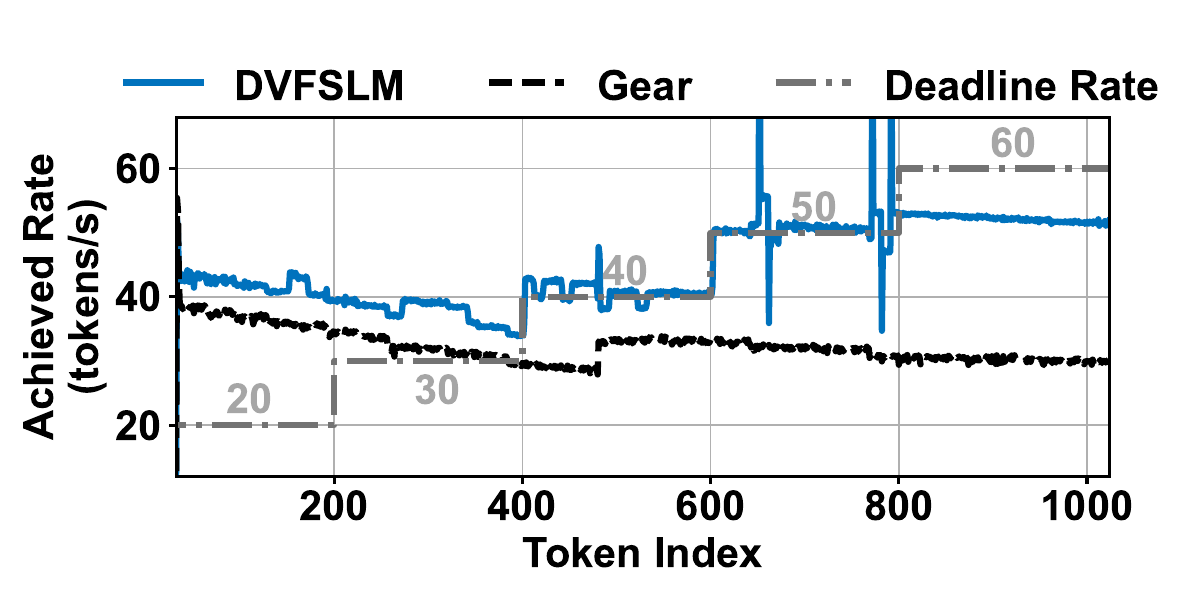}
        \vspace{-.2in}
        \captionof{figure}{Varying the deadline rate at runtime from 20 to 60 tokens/s when running GPT2-large.}
        \label{fig:eval:adjust_target}
    \end{minipage}
    \hfill
    \begin{minipage}[t]{0.33\textwidth}
        \centering
        \includegraphics[width=\linewidth]{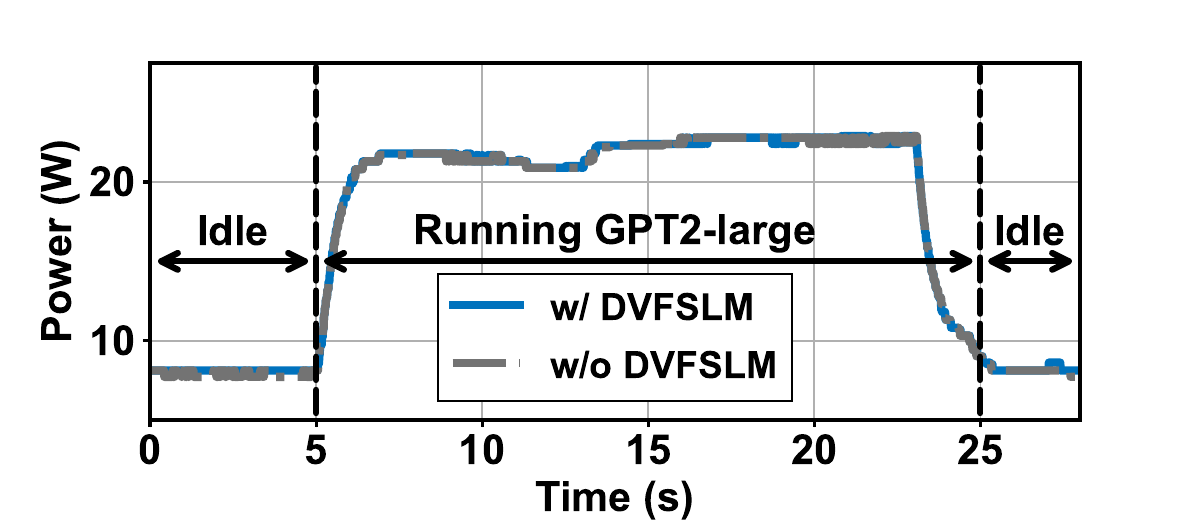}
        \vspace{-.2in}
        \captionof{figure}{Power consumption overhead of \systemname.}
        \label{fig:overhead}
    \end{minipage}
    \vspace{-.15in}
\end{figure*}

\parahead{End-to-end inference} We evaluate an end-to-end inference scenario that includes both prefilling and decoding stages. We vary the prompt length from 100 to 800 tokens with a decoding length up to 1024 tokens, covering commonly used prompt lengths, and run GPT2-large at a decoding deadline rate of 40 tokens/s. Figure~\ref{fig:eval:prefill_energy} and Figure~\ref{fig:eval:prefill_time} report the prefilling and decoding results, respectively. As the prompt length grows, the total energy increases for all methods. In the prefilling stage, \systemname uses high frequencies to minimize the prefilling latency and consumes less energy than Gear and RULE. In the decoding stage, \systemname achieves much lower energy per token than Gear and RULE across different prompt lengths, as it selects frequencies that meet the decoding deadline while minimizing energy. These results show that \systemname remains effective in an end-to-end inference setting.

\parahead{Frequency-decision visualization} To intuitively understand the behavior of each method, we run the decoding stage of GPT2-large at a deadline rate of 20 tokens/s and record the frequency traces in Figure~\ref{fig:eval:freqtrace}. Gear behaves greedily and selects high processor frequencies, resulting in 403.6 mJ per token. \systemname instead co-optimizes the GPU and EMC frequencies, selecting lower frequencies that still meet the deadline while saving energy, yielding only 372.9 mJ per token. RULE adopts conservative GPU frequencies that prolong the decoding latency and miss the deadline, leading to a high 423.4 mJ per token.

\parahead{Thermal behavior} We run different \slms under an ambient temperature of around 25\,$^{\circ}$C, with the device set to the default 30 W power mode and each \slm generating 1024 tokens at a deadline of 40 tokens/s. As shown in Figure~\ref{fig:eval:temp}, the device temperature remains below 43\,$^{\circ}$C across all evaluated \slms, with only minor fluctuations. Although \systemname does not explicitly enforce thermal constraints, its energy-efficient frequency selection helps keep the device temperature low. A stricter deployment can add a constraint such as $T_i\le T_{\max}$ to Eqn.~(\ref{eqn:opt}) or filter frequency candidates by thermal budget.

\parahead{Varying deadline rate at runtime} 
Our system features a configurable token generation rate to adapt to different application or user needs. We verify this capability through an experiment that intentionally changes the rate at runtime. Specifically, we change the deadline rate for \systemname and Gear when running GPT2-large at runtime from 20 to 60 tokens/s for every 200 tokens generated. Through experimental data, we observe that there exists a global minimum energy consumption value for each \slm. This phenomenon occurs because when we increase the frequency from a smaller value, the instantaneous power increases, but the latency decreases. We find that the energy consumption (power multiplied by latency in Eqn.~(\ref{eqn:opt})) leads to a global minimum, and the corresponding token generation rate is not high, \eg, 37, 46, and 43 tokens/s for GPT2-large, Qwen, and TinyLLaMA, respectively.

Therefore, when the deadline rate is low (\eg, 20--30 tokens/s), running GPT2-large at 37 tokens/s is more energy-efficient while also meeting the deadline (\ie, the actual rate achieved is higher than the deadline rate). In Figure~\ref{fig:eval:adjust_target}, \systemname intelligently chooses this higher rate. When the deadline rate is further increased, more token generation will result in more energy consumption. Therefore, \systemname's actual output rate is slightly greater than the required rate, thereby minimizing energy consumption. When the deadline rate reaches 60 tokens/s, we find that \systemname's actual token generation rate is limited by the peak computational capacity of the current device, and can be further improved in the future with more powerful hardware. While Gear's training also seems to find and follow that energy-efficient rate of about 37 tokens/s, it initially runs at this rate, but then fails to cope with deadline rate changes.

\begin{table}[t]
\centering
\setlength{\tabcolsep}{2.5pt}

\caption{Profiling time, storage, and energy of different models. Only about 5\% of the total decoding length is profiled for each model.}
\label{tab:profiling_overhead}
\begin{tabular}{lccc}
\toprule
\textbf{Model} & \textbf{Time (hours)} & \textbf{Storage (GB)} & \textbf{Energy (kWh)} \\
\midrule
GPT2-large & 27.28 & 0.47 & 0.33 \\
Qwen       & 21.37 & 0.34 & 0.25 \\
TinyLLaMA  & 25.12 & 0.61 & 0.31 \\
\bottomrule
\end{tabular}
\end{table}

\subsection{System Overhead}
\label{s:eval:overhead}

\parahead{Profiling overhead} Table~\ref{tab:profiling_overhead} lists the profiling time, storage, and energy for three \slms. The profiling completes within around one day and consumes less than 0.33 kWh per device-model pair, which is acceptable for a one-time calibration. The storage overhead is also small, which is less than 0.61 GB for each model.

\parahead{Runtime overhead} Since we use lightweight latency and power estimators, the time to complete each frequency governing operation of \systemname is short, \ie, 1.73 ms per decision on average, which is over ten times faster than the time needed to generate each token. With the search-acceleration pruning introduced in Section~\ref{s:design:governor}, the decision time (including estimator inference and pruning) is further reduced from 1.73 ms to 0.81 ms, without changing the selected frequencies. Finally, we measure the power overhead of \systemname itself. Figure~\ref{fig:overhead} shows the power measured on AGX Orin. We first fix all processors to the highest frequency, run an \slm, and use its power consumption as the baseline. We then run this \slm again with \systemname with the same settings. The results show that the average power consumption of \systemname increases by 0.2\%, from 20.77 W to 20.81 W (an increase of 0.04 W), which is much lower than the power consumed by running \slm.

\vspace{.12in}
\section{Related Work}
\label{s:related}

\parahead{Power Governing and DVFS} On high-performance platforms (\eg, clouds), workloads are massive (\eg, for LLMs)~\cite{gu2014optimal,kuehn2019dvfs,weng2022mlaas,stojkovic2025tapas} and are processed by a cluster of GPUs~\cite{gu2014optimal,kuehn2019dvfs}. Their power governing focus on adjusting batch size (for inference serving)~\cite{kakolyris2025throttll} and task scheduling among GPUs~\cite{stojkovic2025dynamollm}. Frequency scaling is also performed but at a coarse-grained granularity, such as once every few seconds~\cite{stojkovic2025dynamollm}. Since mobile edge devices have limited computational capabilities and process \slms with a single GPU~\cite{lu2024small}, \slms will dominate the device's workload. Frequency scaling should be performed during \slm execution, typically at sub-second intervals during execution~\cite{kim2021ztt,lin2023workload}, to adapt to workload dynamics. Therefore, existing \dvfs techniques on clouds are unsuitable.

The built-in \dvfs on commercial mobile edge devices uses threshold-based methods with a set of predefined rules~\cite{kim2018survey}, such as nvhost\_podgov for GPU and schedutil for CPU. Although they can adjust frequencies in sub-seconds, their static rules force \slms to operate at fixed token generation rates, which is inefficient for complex workloads~\cite{kim2021ztt}. Existing studies thus propose a series of learning-based DVFS methods~\cite{kim2021ztt,li2022power,geng2024powerlens}. However, they are designed to support traditional deep learning tasks, and do not consider the inherent complexity and unique autoregressive features of \slms, which largely affect their performance for the emerging \slm workloads. CRAVE~\cite{crave2025} is a general-purpose, application-agnostic SoC governor. It analyses the design-induced interaction among CPU, GPU, and main memory with microbenchmarks, and does not model token deadlines or \slm-specific workload. \systemname differs from CRAVE in both target workload and optimization objective. 
Recently, model compression~\cite{liu2025m} and batch size adjustment~\cite{xu2025camel} have been used to jointly optimize energy consumption, latency, and model accuracy for language models running on mobile edge devices. However, these approaches ignore the inherent characteristics of the \slms workload and require a predefined token generation rate, limiting their use in \slm-based applications. In \systemname, we analyze the computation of \slms and propose general, workload-aware power and latency estimators. These estimators guide a runtime DVFS governor that dynamically adjusts processor frequency to minimize energy consumption while meeting target token generation deadlines. Such designs have not been explored in the literature.

\parahead{Resource management of \slms} Although \slms are much smaller than LLMs, they still contain massive parameters~\cite{brown2020language} to ensure performance and consume significant memory and computational resources on mobile edge devices~\cite{lu2024small}. To address these limitations, recent studies have explored various countermeasures. For example, mobile-friendly architectures, \eg, the miniaturized architecture in TinyLLaMA~\cite{zhang2024tinyllama} and the deep and narrow architecture in MobileLLM~\cite{liu2024mobilellm} (since MobiLLM is not supported by llama.cpp, we did not include it in the evaluation), are proposed. Pruning is also used to reduce model size~\cite{fan2025spinfer}. Some methods further slim \slms through on-demand weight loading~\cite{xue2024powerinfer} and weight quantization to adapt to the resource constraints on mobile edge devices. \systemname is orthogonal to these technologies, with weight quantization included in the evaluation, and we plan to study the integration with more such designs in the future.

\parahead{Applications of \slms} Language models serve as a key technology for developing future smart city applications~\cite{shen2025gpiot}. For example, the operation of next-generation controllers or kernels will be significantly automated by language models on embodied robots~\cite{dorbala2023can}, autonomous cars~\cite{sun2024optimizing, han2024pantheon}, and computing devices~\cite{yin2024llm, wang2024swapnet}. Convenient interactions provided by \slms on mobile edge devices will further benefit the fields of healthcare~\cite{yang2024talk2care, ji2025transforming, cao2024practical}, education~\cite{zhang2024simulating}, data analysis~\cite{ma2023demonstration}, and IoT systems~\cite{ouyang2025mmbind, jiang2025freauth, you2026adaptive, li2026longan}. The local deployment and execution of \slms ensures reliability and responsiveness while mitigating the privacy risks associated with data offloading. \systemname complements these benefits by providing energy efficiency under various token generation requirements.

\section{Conclusion}
\label{seq:concl}

This paper introduces \systemname, a new \dvfs design tailored for \slm-driven mobile edge workloads. \systemname advances existing methods along two dimensions: first, it improves energy efficiency by explicitly considering and adapting to complex \slm workloads through novel modeling and design; second, it supports configurable output token generation requirements to be compatible with diverse user or application needs. The requirement does not need to be known in advance and can be adjusted at any time. Extensive evaluations show that \systemname can simultaneously improve both energy efficiency and latency QoS compared to existing methods. A preliminary version of this work has been published in ~\cite{chen2026energy}.

\let\oldbibliography\thebibliography
\renewcommand{\thebibliography}[1]{%
	\oldbibliography{#1}%
	\setlength{\itemsep}{1pt}%
}

\bibliographystyle{IEEEtran}
\bibliography{IEEEabrv,reference}

\end{document}